\documentclass[final,5p,times,twocolumn]{elsarticle}

\usepackage{amsmath,amssymb,amsfonts}
\usepackage{makecell}
\usepackage{graphicx}
\usepackage{siunitx}

\usepackage{lineno, hyperref}
\usepackage{xspace}
\usepackage{subcaption}
\usepackage{gensymb} 
\usepackage{multirow}
\usepackage{placeins}
\usepackage{epsfig}
\usepackage{cuted}

\newcommand{\um}{\si{\micro\metre}}

\makeatletter
\def\ps@pprintTitle{%
	\let\@oddhead\@empty
	\let\@evenhead\@empty
	\def\@oddfoot{}%
	\let\@evenfoot\@oddfoot}

\makeatother
\journal{Nuclear Instruments and Methods A}

\begin{document}
\begin{frontmatter}
\title{Impact of Cold Noise on the tracking performance of ATLAS ITk short strip barrel modules using a charged particle beam}

\author[a]{A.~Affolder}
\author[b]{J.-H.~Arling}
\author[c]{K.~S.~V.~Astrand}
\author[d]{I.~Benaoumeur}
\author[e]{J.~Bucko}
\author[b]{S.~Diez-Cornell}
\author[f]{B.~J.~Gallop}
\author[f]{N.~Ghorbanian}
\author[b]{Y.~He\corref{mycorrespondingauthor}}
\cortext[mycorrespondingauthor]{Corresponding author}
\ead{yajun.he@desy.de}
\author[g]{C.~M.~Helling}
\author[h]{N.~Hessey}
\author[b]{L.~Huth}
\author[i]{C.~Jessiman}
\author[i]{C.~T.~Klein}
\author[i]{J.~S.~Keller}
\author[k]{J.~Kroll}
\author[k]{J.~Kvasnicka}
\author[b]{K.~Mauer}
\author[c]{A.~Murphy}
\author[h,j]{L.~Poley}
\author[h]{D.~M.~Portillo Quintero}
\author[f]{P.~W.~Phillips}
\author[k,l]{R.~Privara}
\author[c,m]{E.~Torres~Reoyo}
\author[k]{P.~Tuma}
\author[n]{M.~Warren}
\affiliation[a]{
    organization={Santa Cruz Institute for Particle Physics, University of California Santa Cruz},
    addressline={1156 High Street},
    city={Santa Cruz, CA 95064},
    country={United States of America}
}

\affiliation[b]{
    organization={Deutsches Elektronen-Synchrotron DESY},
    addressline={Notkestr. 85},
    city={22607 Hamburg},
    country={Germany}
}

\affiliation[c]{
    organization={Fysiska institutionen, Lunds universitet},
    addressline={Sölvegatan 14},
    city={223 62 Lund},
    country={Sweden}
}

\affiliation[d]{
    organization={School of Physics and Astronomy, University of Birmingham},
    addressline={Edgbaston},
    city={Birmingham B15 2TT},
    country={United Kingdom}
}

\affiliation[e]{
    organization={Faculty of Mathematics and Physics, Charles University},
    addressline={Ke Karlovu 3},
    city={121 16 Praha 2},
    country={Czech Republic}
}

\affiliation[f]{
    organization={Particle Physics Department, Rutherford Appleton Laboratory},
    addressline={Harwell Campus},
    city={Didcot OX11 0QX},
    country={United Kingdom}
}

\affiliation[g]{
    organization={University of British Columbia},
    addressline={6224 Agricultural Road},
    city={Vancouver, BC V6T 1Z1},
    country={Canada}
}

\affiliation[h]{
    organization={TRIUMF},
    addressline={4004 Wesbrook Mall},
    city={Vancouver, BC V6T 2A3},
    country={Canada}
}

\affiliation[i]{
    organization={Department of Physics, Carleton University},
    addressline={1125 Colonel By Drive},
    city={Ottawa, ON K1S 5B},
    country={Canada}
}

\affiliation[j]{
    organization={Simon Fraser University},
    addressline={Shrum Science P8429, 8888 University Drive},
    city={Burnaby, BC V5A 1S6},
    country={Canada}
}

\affiliation[k]{
    organization={Institute of Physics of the Czech Academy of Sciences},
    addressline={Na Slovance 2},
    city={182 21 Praha 8},
    country={Czech Republic}
}

\affiliation[l]{
    organization={Faculty of Science, Palacky University},
    addressline={Krízkovského 511/8},
    city={771 47 Olomouc},
    country={Czech Republic}
}

\affiliation[m]{
    organization={Instituto de Física Corpuscular (IFIC), Centro Mixto Universidad de Valencia - CSIC},
    addressline={Catedrático José Beltrán, 2},
    city={46980 Paterna},
    country={Spain}
}

\affiliation[n]{
    organization={Department of Physics and Astronomy, University College London},
    addressline={Gower Street},
    city={London WC1E 6BT},
    country={United Kingdom}
}

\begin{abstract}
The inner tracking system of the ATLAS experiment will be upgraded to a full silicon detector in 2030 for HL-LHC. The new tracking system is called ITk, the Inner Tracker. It is required to be operable with efficiency higher than 99\% and noise hit occupancy smaller than 0.1\%. During the pre-production phase of the ITk project, many short-strip modules were observed to exhibit so-called "Cold Noise (CN)", wherein clusters of strips displayed very high noise when the modules were operated at temperatures below~$-35\degree$C. To investigate the CN impact and ensure the quality of module production, huge amount of effort have been put in by the collaboration. This paper focuses on the impact of CN on the tracking performance by examining two short strip modules that exhibit CN: one is non-irradiated, while the other one has been irradiated to the maximum expected end-of-lifetime fluence. For each module, the global and single strip tracking performance are evaluated. The global performance study shows that the non-irradiated module can be operated within specifications with a threshold around 1fC, but it is not possible to operate the irradiated module as required. In the single strip analysis, it was found that while CN does not affect the charge collection, it reduces the operating window and leaves less margin for detector operation. In the non-irradiated module, less than~3\% of strips fail the detector requirements in the CN regions. For the irradiated module, about~20\% of strips fail the requirements in the low CN region and around~52\% fail in the high CN region. The fraction of strips that cannot operate due to CN throughout their lifetime can be predicted according to the measured~$\mathrm{Q_c^{no}}$ and its expected median collected charge. If the hit noise occupancy caused by CN is kept below~1\% with a threshold greater than 0.45 fC, approximately 60\% of strips could meet the operating requirements by the end of detector's lifetime. Thus, the module is likely to satisfy the operating requirements in terms of global efficiency and global noise occupancy. 
\end{abstract}

\end{frontmatter}

\section{Introduction}
The Inner Tracker (ITk) is a full-silicon tracking detector developed for the ATLAS experiment, specifically designed for the High Luminosity phase of the Large Hadron Collider (HL-LHC). It comprises both pixel and strip detectors. The strip detectors are positioned in the outer layers of the ITk, including the barrel and end-cap regions, ensuring coverage of the entire radius of the solenoid's inner bore. The total coverage area is approximately 165 m$^2$, consisting 17888 modules. Among these modules, 21.3\% are short strip modules, covering the innermost layers of the barrel region.

During the pre-production phase of the ITk project, many short strip modules exhibited clusters of channels with noise level falling outside of the required specifications when tested at operating temperatures of~$-35\degree$C or lower; this issue is referred to as "Cold Noise (CN)". Further investigations revealed that this CN is caused by vibrating capacitors on a flexible printed circuit board (PCB) glued to the sensor surface~\cite{Dyckes_2024}. These vibrations couple into the silicon sensor via the glue and generate unwanted electrical signals.

The ITk strip modules are required to achieve tracking efficiency higher than 99\% and to keep hit noise occupancy lower than 0.1\% throughout their operational lifetime. This paper aims to highlight the impact of CN on tracking performance and to determine whether the modules can operate within specifications when affected by CN. The input data for the analyses in this paper are from two short strip modules with identified CN: one is non-irradiated, while the other one has been sensor-irradiated to a fluence of $1.1\times10^{15}$ 1 MeV $\mathrm{n_{eq}/cm^{2}}$ using 24 GeV protons from the IRRAD facility~\cite{ravotti:hal-03011049}. This corresponds to the maximum expected end-of-lifetime fluence for any short strip module, including a 50\% safety factor. CN modules have different levels and patterns of noise. Therefore, in this paper, both the single strip performance as a function of its noise and the global performance are studied, so that the results can be extrapolated to other modules.

\section{ITk Short Strip Module}
\label{sec:ss}
\begin{figure}
\centering
\includegraphics[width=1.0\linewidth]{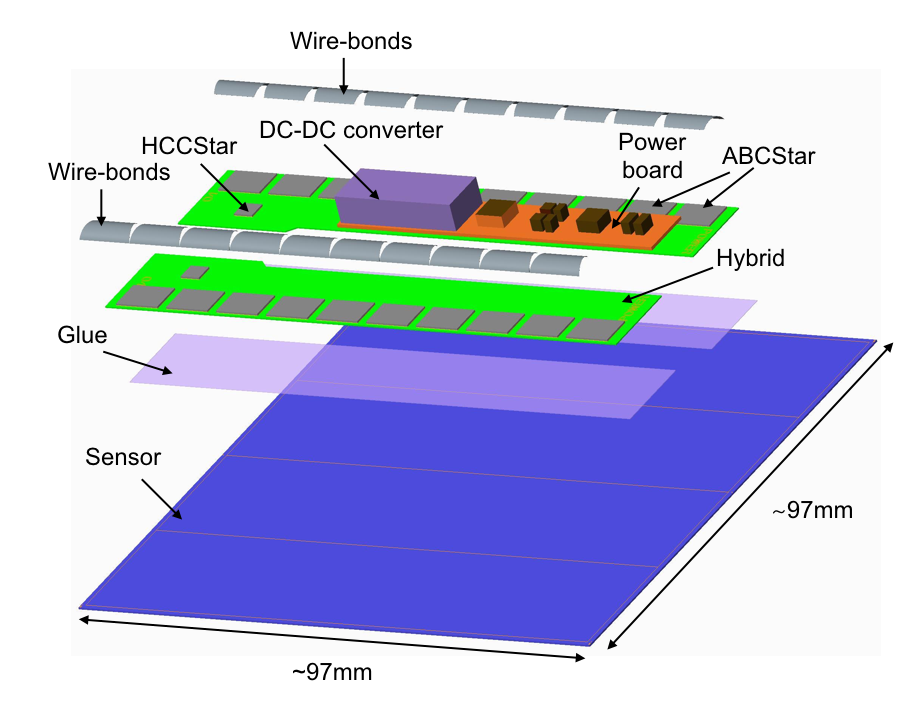}
\caption{Exploded view of a short strip barrel module with all relevant components. Taken from~\cite{CERN-LHCC-2017-005}.}
\label{module}
\end{figure}

A short strip module~\cite{CERN-LHCC-2017-005} consists of one sensor, two low-mass PCBs known as hybrids that house the readout and control ASICs — the ABCStar and HCCStar — as well as a power board. The power board features a DC-DC converter, a GanFET transistor, and a monitoring and control ASIC (AMACv2). An exploded view of a short strip module is shown in Figure~\ref{module}. The sensor of the ITk short strip module is of the n$^+$-in-p type, produced from a 6-inch float-zone p-type silicon wafer~\cite{Unno_2023}. It is divided into four segments, with 1282 parallel strips per segment, which includes 1280 readout strips and one dummy strips on each side to help shape the electric field. Each strip has a pitch of 75.5~\um \ and a length of about~2.5~cm. The flex hybrids and the power board are glued directly to the silicon sensor using electronics-grade epoxy. 

\begin{figure*}
\centering
	\begin{subfigure}{0.49\linewidth}
		\includegraphics[width=\linewidth]{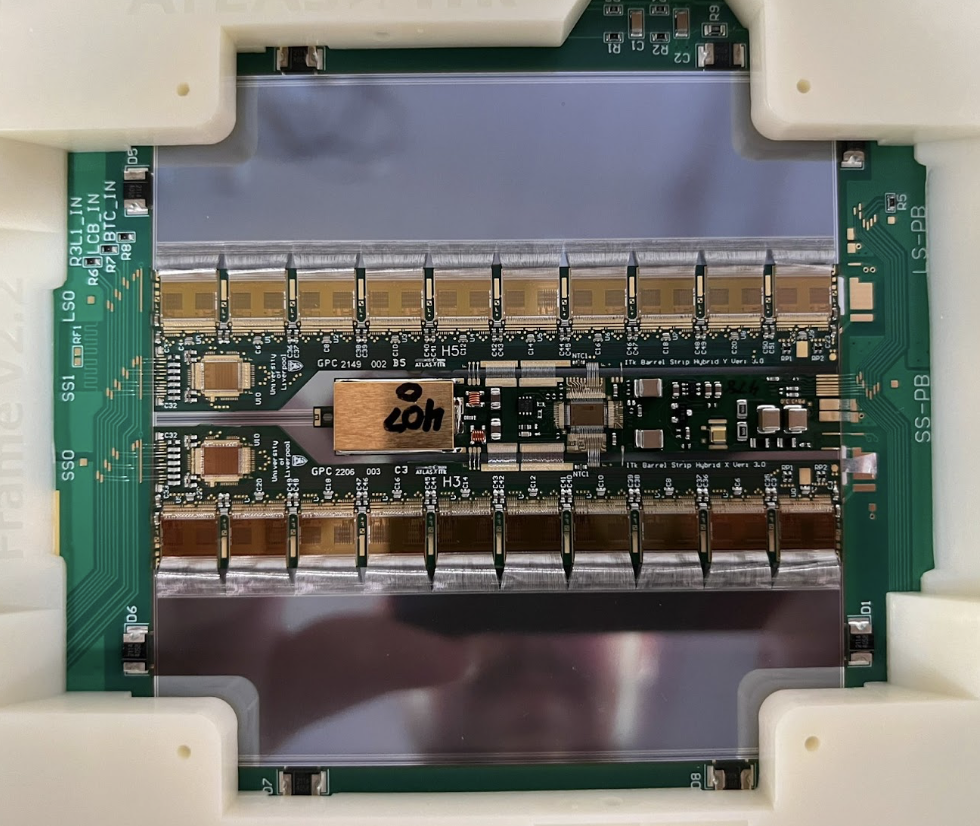}
		\caption{Module}
        \label{fig:module}
	\end{subfigure}	
	\begin{subfigure}{0.49\linewidth}
		\includegraphics[width=\linewidth]{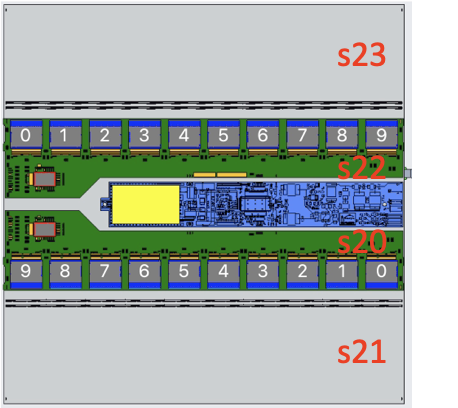}
		\caption{Numbering scheme}
        \label{fig:sconvention}
	\end{subfigure}	
\caption{(a) A physical short strip module attached to its test frame. (b) Numbering scheme for the SS module, with the DC-DC converter highlighted in a yellow rectangle.}
\label{fig:physical-module}
\end{figure*}

Figure~\ref{fig:module} displays a short strip module mounted on its test frame. In the picture, the hybrids glued onto the inner two segments of the sensor and the power board is glued between the hybrids. Each hybrid holds 10 ABCStar read-out chips~\cite{Lu_2017, Cormier_2021}, with each ABCStar chip providing binary output for 256 channels. Half of the channels come from the segment beneath the hybrid, and the other half come from the segment away from the hybrid. Figure~\ref{fig:sconvention} illustrates the numbering scheme used in this paper. The hybrid on the lower half of the module communicates with the lower two segments, labeled s20 and s21. The ABCStar chips, arranged from right to left, are labeled from 0 to 9. Similarly, the active strips in both s20 and s21 are numbered from 0 to 1279 from right to left. The second hybrid, positioned at the upper part of the module, communicates with the upper two segments, labeled s22 and s23. For these chips and strips, the numbering scheme follows a left-to-right order. Vibrating components originate from the DC-DC converter, and CN only occurs in the sensor segments overlapping with the power board, i.e. s20 and s22, the investigation in this paper focus on these segments, with the other two segments for reference.

\section{Noise Estimation}
\label{sec:no}
ABCStar chips utilize a binary readout scheme. The threshold scan is the most common way by which the noise of a binary detector system is measured. At each threshold level, noise hits are recorded with a given external trigger rate. The logarithm of hit occupancy due to Gaussian noise is proportional to the square of the threshold \cite{Spieler:2005si}. 
\begin{equation}
    \mathrm{P_n = \frac{\Delta t}{4\sqrt{3} \tau} \cdot exp(-Q_T^2/2Q_n^2)}
\end{equation}
$\mathrm{P_n}$ is the noise hit occupancy in a time interval~$\mathrm{\Delta t}$.~$\tau$ is the time constant of an RC low-pass filter.~$\mathrm{Q_T}$ is the threshold level and~$\mathrm{Q_n}$ is the equivalent noise charge at the input of the pre-amplifier. The value of~$\mathrm{Q_n}$ which can be derived by the slope of~$\mathrm{log (P_n)}$ versus~$\mathrm{Q_T^2}$.

The nature of CN, i.e how the mechanical effect couples to the electrical effect, has not been fully revealed, making it challenging to accurately characterize the equivalent noise levels for strips affected by CN. Details can be found in Figure~\ref{fig:pedestal-no-curve}. As a result, this paper introduces~$\mathrm{Q_c^{no}}$ to quantify the noise behavior of each strip.~$\mathrm{Q_c^{no}}$ corresponds to the threshold for which the hit noise occupancy is~$\mathrm{10^{-3}}$, as the strip modules of the ITk are required to achieve a noise hit occupancy of less than 0.1\% to see pile-up events and to ensure reliable and accurate tracking.~$\mathrm{Q_c^{no}}$ is obtained from the fitted $\mathrm{log(P_n)}$ versus~$\mathrm{Q_T^2}$ curve using the threshold region of~$\mathrm{P_n < 10^{-2}}$. Figures~\ref{fig:pedestal-noise-unirr} and~\ref{fig:pedestal-noise-irr} display the value of~$\mathrm{Q_c^{no}}$ for the non-irradiated and irradiated modules, respectively.

In this paper, we specifically identify the regions referred to as the CN regions as follows:
\begin{itemize}
\item non-irradiated, s20, strips 640 -1023, high CN
\item non-irradiated, s22, strips 0 - 255, low CN
\item non-irradiated, s22, strips 256 - 511, high CN
\item Irradiated, s20, strips 640 - 1023, low CN
\item Irradiated, s22, strips 256 - 639, high CN
\end{itemize}

The strips not listed above fall within the normal strip regions. The average~$\mathrm{Q_c^{no}}$ measured for normal strips of s21 and s23 is~0.31~$\pm$~0.02~fC for both non-irradiated and irradiated modules. The average~$\mathrm{Q_c^{no}}$ of strips which are not from the CN regions of s20 and s22 is between~0.32~fC and~0.33~fC.~$\mathrm{Q_c^{no}}$ for strips from the CN region is generally above the average of~$\mathrm{Q_c^{no}}$, with some strips reaching up to~1.3~fC. Different CN characteristics of non-irradiated and irradiated modules are not caused by the irradiation itself, but rather by the variation of the CN behavior among individual modules and particular testing conditions.

\section{Test Beam Setup}
\label{sec:setup}
The modules were tested at the DESY II testbeam facility~\cite{DIENER2019265} with 5 GeV/c electrons. The ADENIUM telescope~\cite{Liu:2023uas} was employed to reconstruct reference tracks. The intrinsic resolution of the ADENIUM sensor is estimated to be 8.44~\um~in the x-dimension and 7.76~\um~in the y-dimension, which are better than the short strip module. The module under test was mounted in a cooling box that achieved an ambient air temperature of~$-40\degree$C; the box was placed in the middle of the telescope. A scintillator located in front of the telescope generated a signal each time a particle passed through, which was used to feed the AIDA-2020 Trigger Logic Unit~\cite{Baesso:2019smg} for trigger generation. It is important to note that ADENIUM telescope is not able to resolve multiple hits recorded in the same trigger. Therefore, behind the telescope, \texttt{TelePix2} pixel detector~\cite{HUTH2025170720} was used as a timing plane to resolve these ambiguities. The EUDAQ2 framework managed the data collection from all detectors~\cite{Liu:2019wim}, while data reconstruction and analysis of the collected data were carried out using the Corryvreckan framework~\cite{Dannheim:2020jlk}.

\begin{figure*}
\centering
	\begin{subfigure}{0.49\linewidth}
		\includegraphics[width=\linewidth]{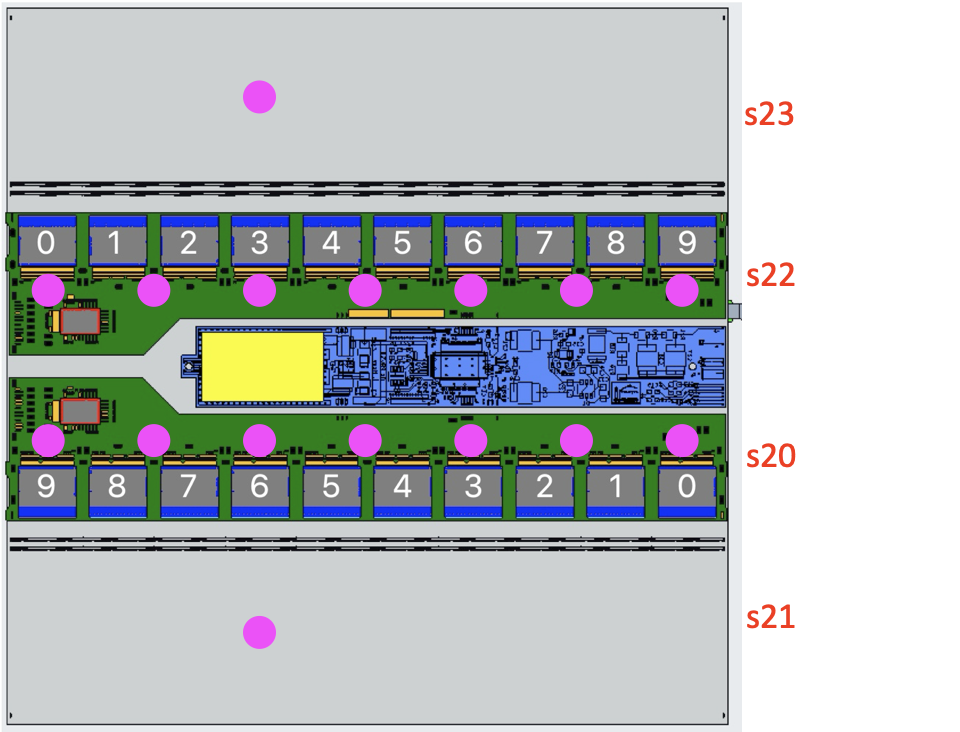}
		\caption{non-irradiated module}
        \label{fig:unirr-beam-pos}
	\end{subfigure}	
	\begin{subfigure}{0.49\linewidth}
		\includegraphics[width=\linewidth]{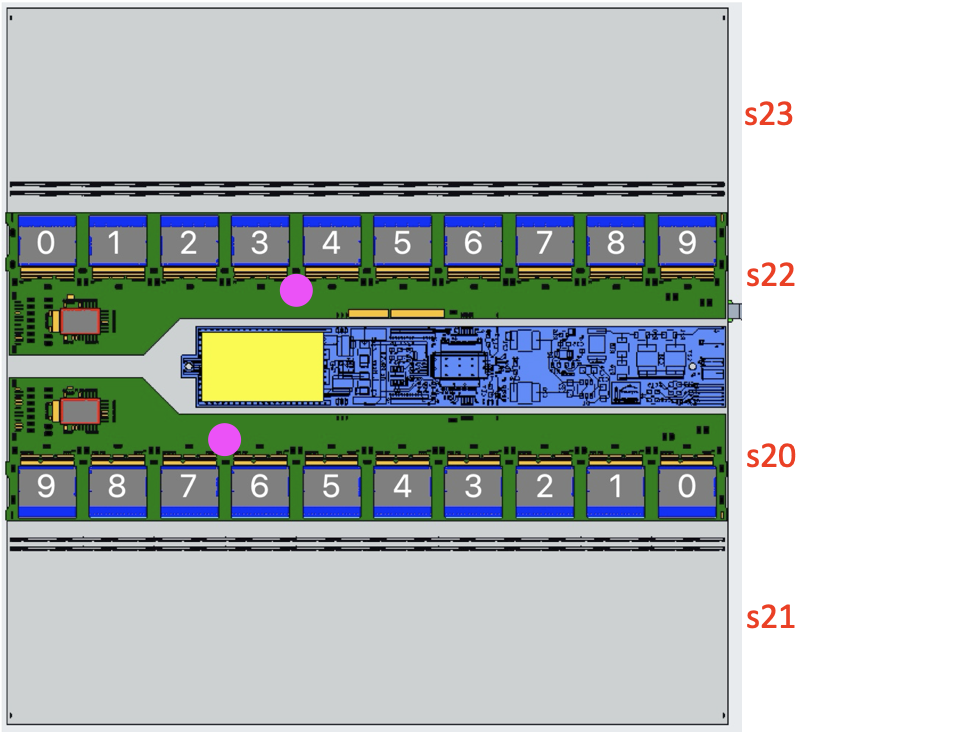}
		\caption{Irradiated module}
        \label{fig:irr-beam-pos}
	\end{subfigure}	
\caption{Positions selected for threshold scans of both the non-irradiated and irradiated modules. The purple dots indicate the center of the beam spot.}
\label{fig:beam-pos}
\end{figure*}

To assess the efficiency of all strips in s20 and s22 of the non-irradiated module, threshold scans were conducted at seven different positions at each stream. A single threshold scan was performed in s21 and s23, serving as a reference for the non-irradiated module. For the irradiated module, threshold scans were carried out in the CN regions of streams 20 and 22 to analyze the impact of CN on tracking, which overlaps with the effects of irradiation. Figure~\ref{fig:beam-pos} shows the center of the beam position for both the non-irradiated and irradiated modules.

\section{Single Strip Performance}
\label{sec:eff_ana}
\subsection{Efficiency Analysis}
\begin{figure*}
    \centering
    \begin{subfigure}{0.49\linewidth}
        \includegraphics[width=\linewidth]{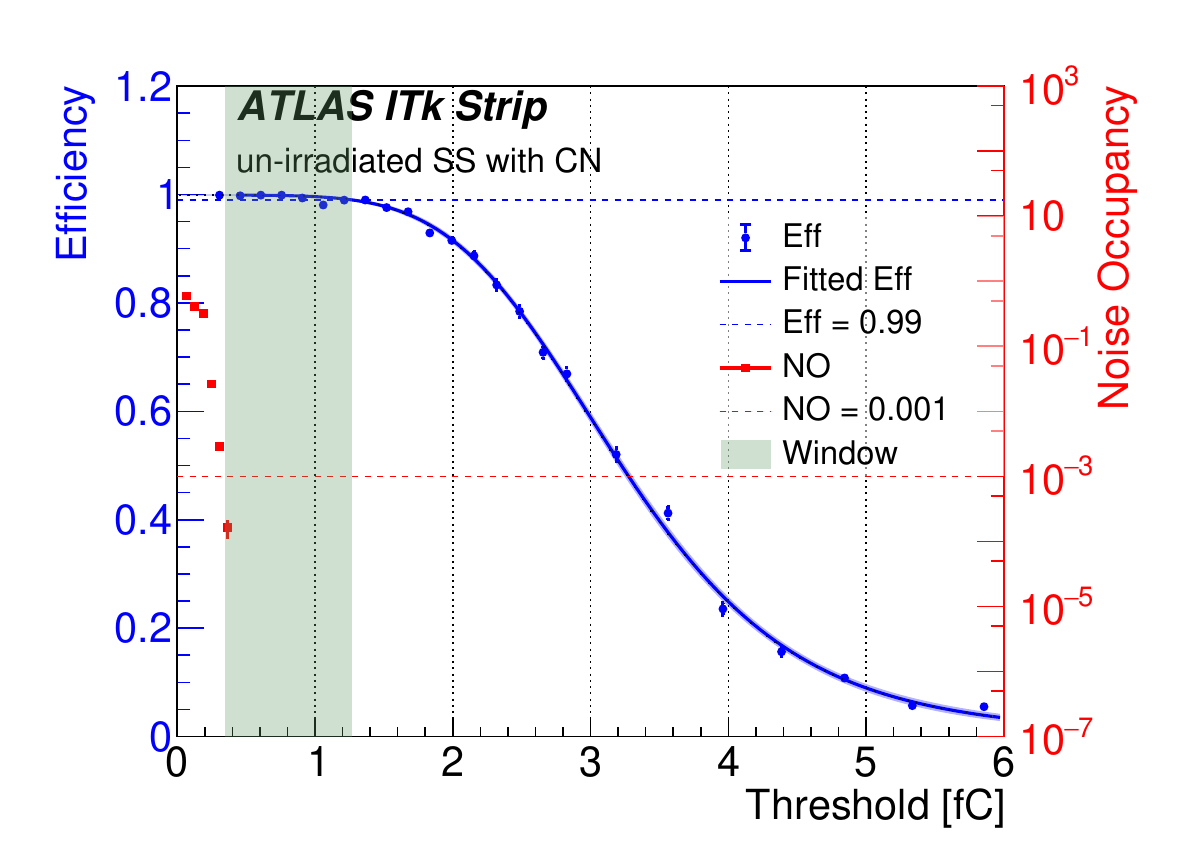}
        \caption{Normal strip, non-irradiated}
        \label{fig:unirr-normal-eff}
    \end{subfigure}
    \begin{subfigure}{0.49\linewidth}
        \includegraphics[width=\linewidth]{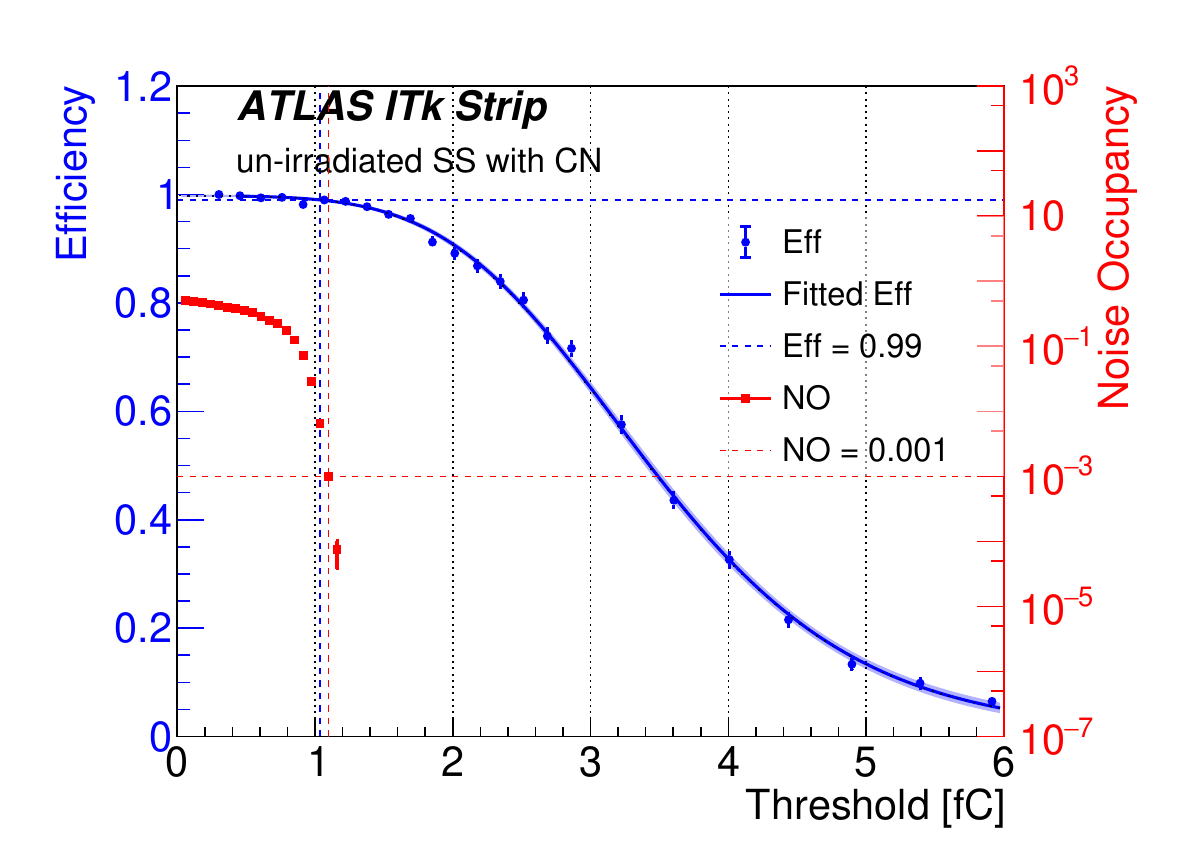}
        \caption{CN strip, non-irradiated}
        \label{fig:unirr-noise-eff}
    \end{subfigure}
    \begin{subfigure}{0.49\linewidth}
        \includegraphics[width=\linewidth]{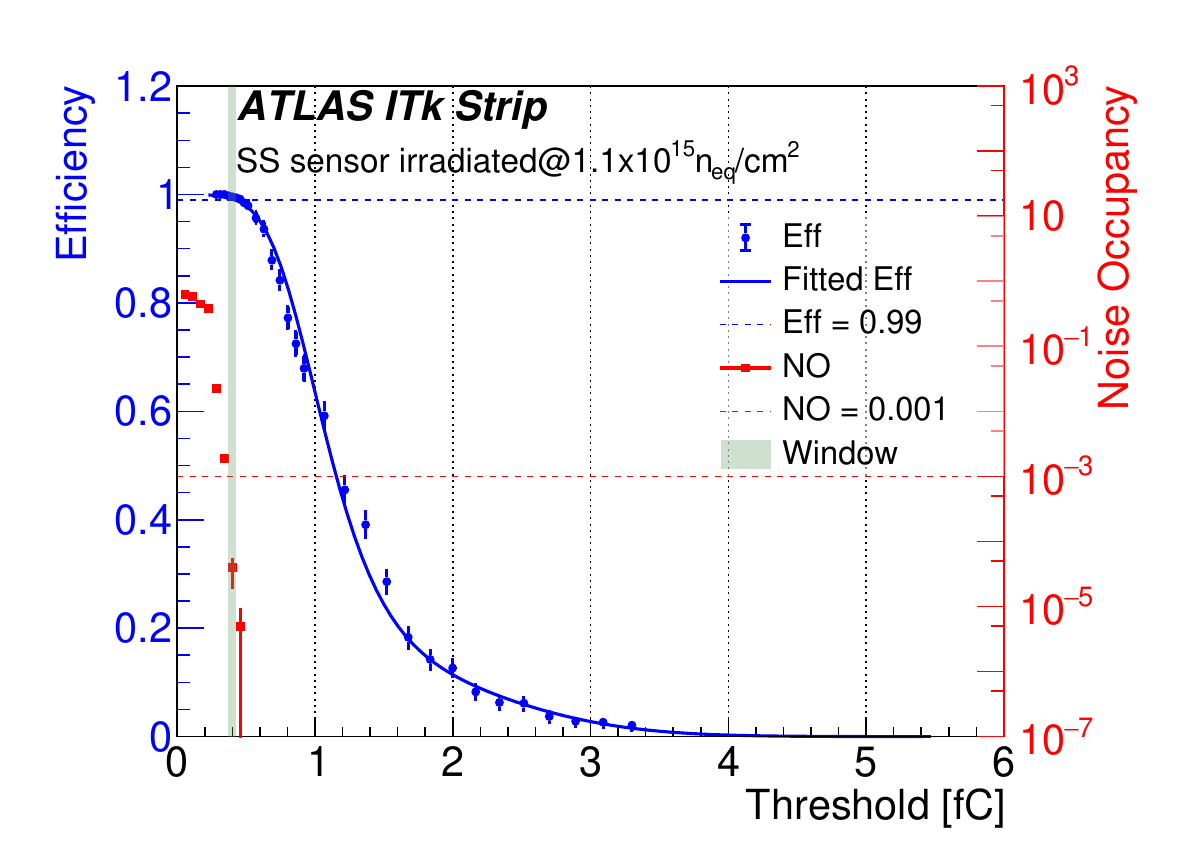}
        \caption{Normal strip, irradiated}
        \label{fig:irr-normal-eff}
    \end{subfigure}
    \begin{subfigure}{0.49\linewidth}
        \includegraphics[width=\linewidth]{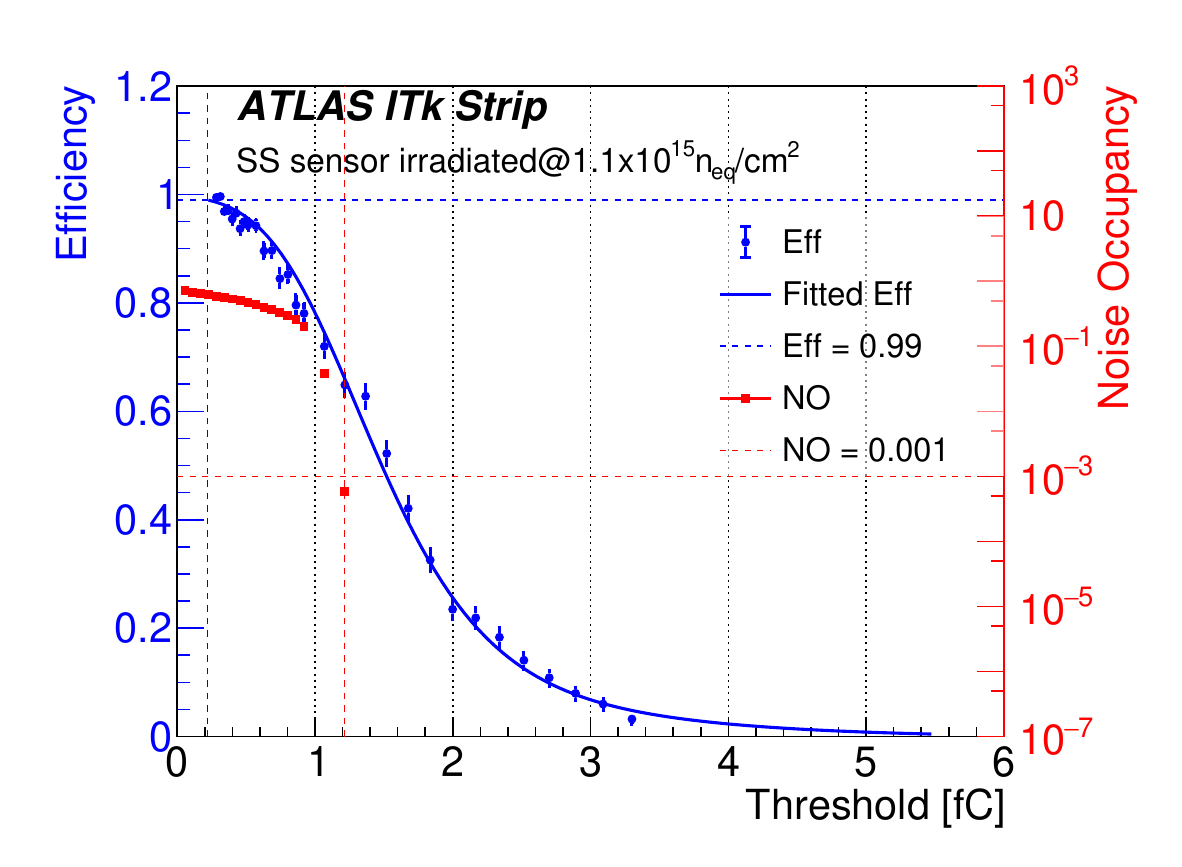}
        \caption{CN strip, irradiated}
        \label{fig:irr-noise-eff}
    \end{subfigure}
    \caption{The detection efficiency of selected single strips (shown in blue) and noise occupancy (shown in red) are plotted as a function of the charge threshold for a normal strip and a strip affected by CN in both non-irradiated and irradiated modules. The efficiency points are fitted using Equation~\ref{eq:eff_fit} up to 5.0 fC and extended to 6.0 fC for the non-irradiated module. The dashed blue line indicates the required efficiency level, while the dashed red line represents the required noise occupancy level. The pale green region illustrates the operating window if~$\mathrm{Q_c^{no}<Q_c^{eff}}$.}
    \label{fig:eff}
\end{figure*}

The efficiency of a single strip for a specific threshold was determined by calculating the ratio of the number of tracks with associated hits to the total number of tracks that intersect that strip. To model the efficiency points as a function of the threshold, an empirical skewed complementary error function was utilized

\begin{equation}
    f(x)=\frac{1}{2}\cdot\epsilon_\text{max}\cdot\text{erfc}\left[\frac{x-Q_{50}}{\sqrt{2}\sigma}\cdot\left(1-0.6\tanh\left(\frac{A(x-Q_{50})}{\sqrt{2}\sigma}\right)\right)\right]\,
    \label{eq:eff_fit}
\end{equation}

where~$\text{erfc}$ denotes the complementary error function,~$\tanh$ is the hyperbolic tangent function,~$\epsilon_\text{max}$ is the maximum efficiency,~$\sigma$ is the width of the error function distribution, and~$\mathrm{A}$ is a skew parameter~\cite{eff_fit-ref, Arling:2023pio, Arling_2025}. The parameter~$\mathrm{Q_{50}}$ represents the charge threshold at which exactly~50\% of the tracks register a hit in the modules under test. This corresponds to the median of the charge distribution collected on the leading strip. Figure~\ref{fig:eff} illustrates the efficiency as a function of the charge threshold for both a normal strip and a CN strip, in both the non-irradiated and irradiated modules. The function given in the equation can accurately model the efficiency curve in all cases.

The fitted curves are primarily used for two analyses in this paper: estimation of the median collected charge, denoted as~$\mathrm{Q_{50}}$, and the width of operating window. 

\subsection{Median Collected Charge}

\begin{table}[]
    \centering
    \begin{tabular}{c|c|c}
    \hline
        Stream& non-irradiated & Irradiated \\
        \hline
        s20 & 3.37 $\pm$ 0.09 fC& 1.26 $\pm$ 0.04 fC\\
        s21 & 3.32 $\pm$ 0.07 fC& - \\
        s22 & 3.40 $\pm$ 0.09 fC& 1.18 $\pm$ 0.08 fC\\
        s23 & 3.33 $\pm$ 0.08 fC& - \\
        \hline
    \end{tabular}
    \caption{Average of~$\mathrm{Q_{50}}$ for each stream. When calculating the average of~$\mathrm{Q_{50}}$, the following strips are excluded: dead strips and the strips of threshold above 2 fC are not scanned.}
    \label{tab:q50}
\end{table}

The median collected charge for each strip in all streams of the non-irradiated and irradiated module are displayed in ~\ref{app:q50}. The results indicate that charge collection is not significantly impacted by CN. The average median collected charge for the non-irradiated and irradiated modules is summarised in Table~\ref{tab:q50}. One should note that the median collected charge of the irradiated module is less than~40\% of that of the non-irradiated module due to the bulk defects resulting from irradiation. 

\subsection{Operating Window}

\begin{table*}[]
    \centering
    \begin{tabular}{c|p{5em}|p{2em}|p{2em}|p{9em}|p{9em}}
    \hline
        Module & Region & Total strips & Dead strips & Failed strips near a dead strip (boundary) & Failed strips due to CN\\
         \hline
         \multirow{7}{6em}{non-irradiated} 
         &s20 normal & 896 & 1 & 2 (2) & 0 \\
         &s20 high CN & 384 & 5 &  8 & 1 \\
         &s21 normal& 351 & 2 &  4 & 0 \\
         &s22 normal& 768 & 0 &  0 (1) & 0\\
         &s22 low CN& 128 & 1 &  2 (1) & 0\\
         &s22 high CN& 256 & 3 &  4 & 8 \\
         &s23 normal& 340 & 1 &  2 & 0 \\
         \hline
         \multirow{2}{6em}{Irradiated}&s20 low CN & 256 & 0 & 0 & 51 \\
         &s22 high CN & 261 & 0 & 0 & 137 \\
         \hline
    \end{tabular}
    \caption{The number of total strips, the number of dead strips, the number of failed strips due to a dead strip or at the boundary region of the sensor and the number of failed strips due to CN in each region. The total number of strips covered by position scans is listed in the third column. Dead strips are strips with~$\mathrm{Q_c^{no}=0}$. Failed strips are active strips with~$\mathrm{Q_c^{no} \geq Q_c^{eff}}$.}
    \label{tab:strips}
\end{table*}

The operating window is defined as the threshold region where the strip achieves an efficiency greater than the required level, while the noise hit occupancy remains below the specified threshold. The strip modules of the ITk are required to operate with an efficiency exceeding~99\%\footnote{Since~\texttt{TelePix2} data were not available for the scans of the non-irradiated module, according to~\cite{HUTH2025170720}, for position scans on strip 768-1280 of s22, ~$\mathrm{Q_c^{eff}}$ is approximated by the threshold where the efficiency is~98\%.}. The highest threshold satisfying this requirement is referred to as~$\mathrm{Q_c^{eff}}$ and is obtained using the fitted curves. Additionally, since the strip modules must have noise hit occupancy lower than~0.1\%, together with the value of~$\mathrm{Q_c^{no}}$ calculated in Section~\ref{sec:no}, the operating window can be derived if~$\mathrm{Q_c^{no}<Q_c^{eff}}$. 

In the CN regions of s20 and s22,~$\mathrm{Q_c^{eff}}$ is reduced depending on~$\mathrm{Q_c^{no}}$. When the noise level of a strip increases, the width of the error function distribution, as detailed in Equation~\ref{eq:eff_fit}, becomes larger. Consequently, the high-efficiency plateau in the efficiency versus threshold curve can easily fall below~99\% for CN affected strips, as depicted in Figure~\ref{fig:unirr-noise-eff} and~\ref{fig:irr-noise-eff}. Detailed results of ~$\mathrm{Q_c^{eff}}$ for the non-irradiated and irradiated modules are attached in~\ref{app:opw}. The results show that, before irradiation, an operating window of more than 1 fC is identified for normal strips; following irradiation, an operating window of approximately 0.1 fC is identified. 

A strip with~$\mathrm{Q_c^{no} \geq Q_c^{eff}}$ fails the requirement and no operating window can be identified. In this study, there are two main reasons of a failure: a strip fails due to a dead strip and a strip fails due to high noise. Around a dead strip, especially at low threshold, the hit cluster position is mis-estimated due to a missing hit from the dead strip. The cluster-track association gets more likely missed and the measured efficiency gets lower. Same reason for the strips at the boundary of sensor. Table~\ref{tab:strips} summarises the number of dead strips, the number of failed strips due to a dead strip or at the boundary and the number of failed strips due to noise. For the non-irradiated module, no strips fails in the normal and low CN region and~ less than 3\% of strips fail in the high CN regions. For the irradiated module, around~20\% of strips fail in the low CN region and around~52\% of strips fail in the high CN region.

\begin{figure*}[]
    \centering
    \includegraphics[width=0.6\linewidth]{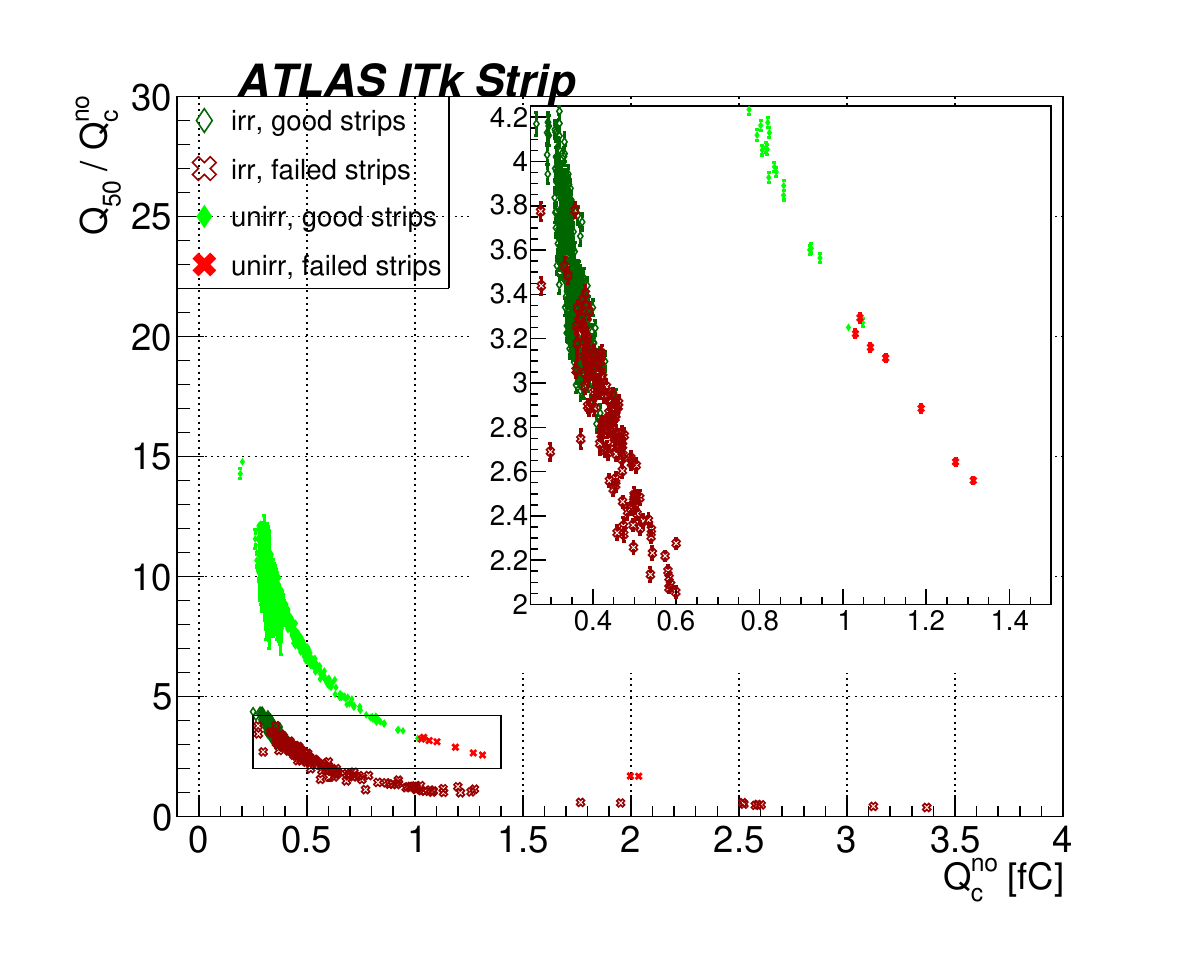}
    \caption{The relationship between of $\mathrm{Q_{50}}$/$\mathrm{Q_c^{no}}$ and~$\mathrm{Q_c^{no}}$ using data from strips of the non-irradiated and irradiated modules. The dead strips and their neighbour strips are excluded. Good strips are strips with~$\mathrm{Q_c^{eff}>Q_c^{no}}$. Failed strips are strips with~$\mathrm{Q_c^{eff} \leq Q_c^{no}}$.}
    \label{fig:sumsn}
\end{figure*}

\begin{figure*}[]
    \centering
    \includegraphics[width=0.6\linewidth]{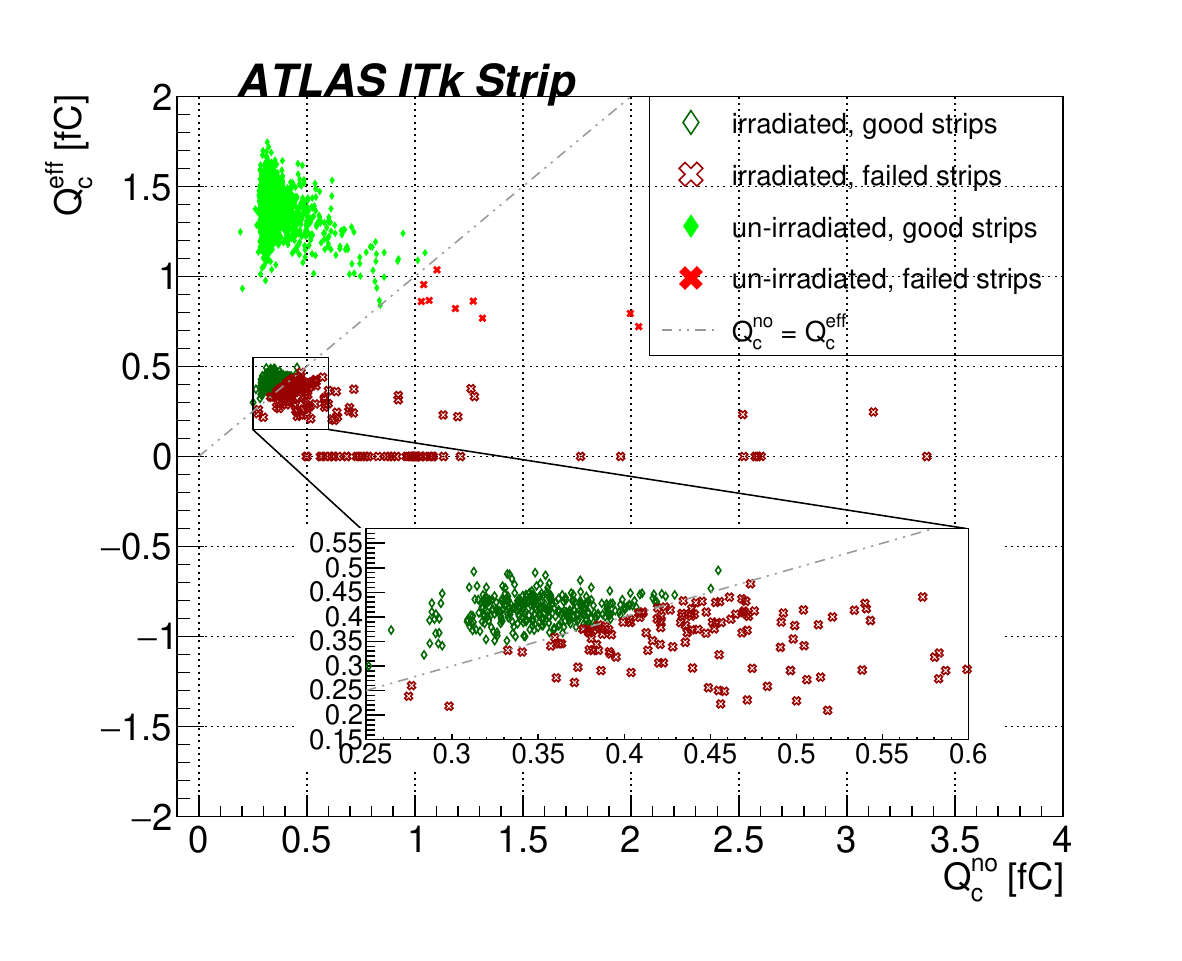}
    \caption{The relationship between~$\mathrm{Q_c^{eff}}$ and~$\mathrm{Q_c^{no}}$ using data from strips of the non-irradiated and irradiated modules. The dead strips and their neighbour strips are excluded. Good strips are strips with~$\mathrm{Q_c^{eff}>Q_c^{no}}$. Failed strips are strips with~$\mathrm{Q_c^{eff} \leq Q_c^{no}}$.}
    \label{fig:sumeff}
\end{figure*}

Figure~\ref{fig:sumsn} summarises the relationship between the ratio of~$\mathrm{Q_{50}/Q_c^{no}}$ and~$\mathrm{Q_c^{no}}$ using data from strips of the non-irradiated and irradiated modules. The failed strips and good strips are highlighted with different colors. Before irradiation, strips with ~$\mathrm{Q_{50}/Q_c^{no}> 3.3}$ meet the operating requirement and strips with ~$\mathrm{Q_{50}/Q_c^{no}< 3.3}$ are unlikely to have an operating window. After irradiation, 
\begin{itemize}
    \item With ~$\mathrm{Q_{50}/Q_c^{no}> 3.3}$, up to 96\% of strips meet the operating requirement.
    \item  With ~$\mathrm{2.8 < Q_{50}/Q_c^{no}<3.3}$, about 64\% of strips have an operating window.
    \item With ~$\mathrm{Q_{50}/Q_c^{no}< 2.8}$, no strip has operating window.
\end{itemize}
~$\mathrm{Q_{50}}$ reduces with the detector operation time as the irradiation fluence goes up. Therefore, the fraction of failed strips due to CN can be predicted according to the relationship of~$\mathrm{Q_{50}/Q_c^{no}}$ and~$\mathrm{Q_c^{no}}$. It is observed that in both non-irradiated and irradiated cases, a value of ~$\mathrm{Q_{50}/Q_c^{no}}>3.3$ is necessary to have a high likelihood of passing the requirements. For each module, ~$\mathrm{Q_{50}}$ will decrease with operation time as the irradiation fluence accumulates, in a predictable way which depends on the module location in the detector. Therefore, the fraction of strips without an operating window due to CN at any point in the lifetime can be predicted according to the measured~$\mathrm{Q_c^{no}}$ on a module and its expected~$\mathrm{Q_{50}}$.

Figure~\ref{fig:sumeff} summarises the relationship between~$\mathrm{Q_c^{eff}}$ and~$\mathrm{Q_c^{no}}$. It is observed that~$\mathrm{Q_c^{eff}}$ decreases as~$\mathrm{Q_c^{no}}$ increases, which leads to a reduction of the operating window width. Before irradiation,~$\mathrm{Q_c^{eff}}$ is not higher than 1.8 fC.~$\mathrm{Q_c^{no}}$ must be smaller than 1 fC to get an operating window. With irradiation up to the end-of-life fluence,~$\mathrm{Q_c^{eff}}$ is not higher than 0.5 fC. And
\begin{itemize}
    \item With~$\mathrm{Q_c^{no}> 0.45 \ fC}$, no strip has operating window.
    \item With~$\mathrm{Q_c^{no}< 0.38 \ fC}$, which is the average ~$\mathrm{Q_c^{no}}$ plus the estimated uncertainty, up to~94\% of strips meet the requirements.
    \item With~$\mathrm{0.38 <Q_c^{no}< 0.45 \ fC}$,~45\% of strips have operating window, which corresponds about 63\% of strips if $\mathrm{Q_c^{no}< 0.45 \ fC}$ required.
\end{itemize}

\begin{table*}[]
    \centering
    \begin{tabular}{c|p{12em}|p{8em}|p{8em}}
    \hline
         ~$\mathrm{Q_c^{max}}$ (fC)& Number of masked CN strips & Global~$\mathrm{Q_c^{no}}$ (fC) & Global~$\mathrm{Q_c^{eff}}$ (fC)\\
         \hline
         - & 0 & 1.18 & 0.39 \\
         1.2 & 12 & 0.88 & 0.40 \\ 
         1.0 & 24 & 0.71 & 0.40 \\
         0.8 & 36 & 0.53 & 0.40 \\
         0.6 & 56 & 0.43 & 0.41 \\
         0.45 & 90 & 0.36 & 0.41 \\
         0.4 & 114 & 0.35 & 0.41 \\
         \hline
    \end{tabular}
    \caption{Number of masked CN strips, global~$\mathrm{Q_c^{no}}$ and global~$\mathrm{Q_c^{eff}}$ for different ~$\mathrm{Q_c^{max}}$.}
    \label{tab:mask}
\end{table*}

\section{Global Performance}
At the beginning of the detector's lifetime, the threshold can set up to 1 fC, so that, without masking any noisy strips, the detector can function at 99\% efficiency, with noise occupancy less than 0.1\%. However, as the detector approaches the end of its operational lifetime, due to irradiation, the threshold must be chosen carefully to optimize the overall performance of the module. 

Let's focus on the s22 of the irradiated module and estimate how the CN affects its global tracking performance. The efficiency of strips 0 to 354 and strips 616 to 1279, which were not exposed to the electron beam, is approximated by the average efficiency of strips 550 to 559.  In addition, if~$\mathrm{Q_c^{no}}$ of a CN strip exceeds a maximum allowed value ~$\mathrm{Q_c^{max}}$, the strip will be masked. Table~\ref{tab:mask} summarizes the number of masked CN strips, as well as global~$\mathrm{Q_c^{no}}$ and global~$\mathrm{Q_c^{eff}}$ for different values of~$\mathrm{Q_c^{max}}$. The table shows that the global~$\mathrm{Q_c^{eff}}$ does not significantly depend on the number of masked strips; however, the global ~$\mathrm{Q_c^{no}}$ does. It suggests that the module should operate at a threshold between 0.36 and 0.41 fC, masking 90 CN strips with~$\mathrm{Q_c^{no}}$ greater than  0.45 fC to meet the operating requirements. It is important to note that the module specification also requires that no more than 1\% of the channels in any module or row can be classified as bad. Therefore, it may not be possible to operate this module at the end of its lifetime while remaining within specification.

\begin{figure}[]
    \centering
    \includegraphics[width=1.0\linewidth]{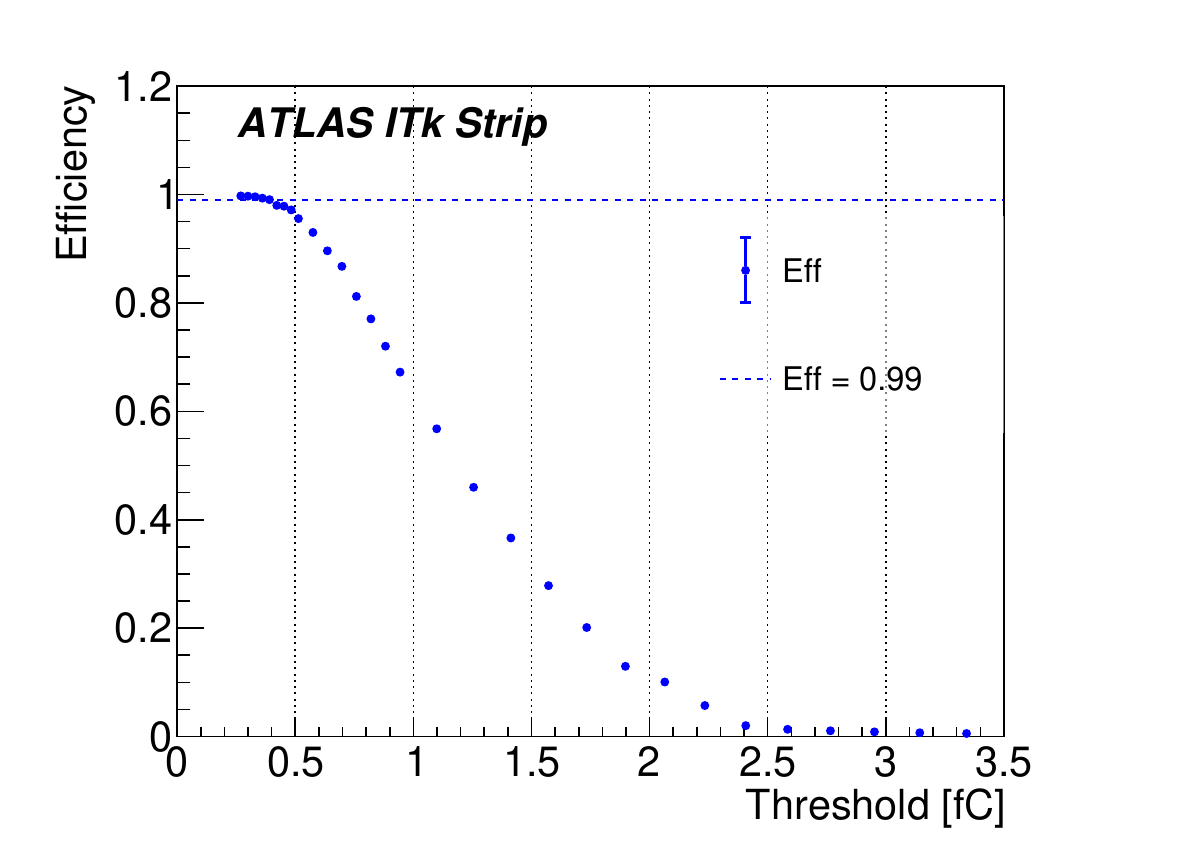}
    \caption{Estimated global efficiency of the irradiated s22 as the function of global threshold. The plots can serve as the reference efficiency plot when deciding the operating threshold.}
    \label{fig:effref}
\end{figure}

Nevertheless, the following operational strategy can be considered: first, a threshold setting of approximately 0.40 fC should be tested, which allows the module to achieve around 99\% efficiency. Ideally, noisy strips should be masked based on their noise occupancy levels until the global noise occupancy is below 0.1\%. If a large number of strips need to be masked, it may be necessary to increase the threshold setting, even at the cost of some global efficiency. Figure~\ref{fig:effref} presents a reference plot of efficiency as the function of the threshold. It is worth noting that the module studied in this paper was irradiated to the maximum fluence for short strip modules. In areas of the barrel region with low fluence, the CN modules may still be usable. Further dedicated studies with CN modules irradiated to varying levels should be conducted. 

\section{Summary and Discussions}

The ITk strip modules are required to achieve tracking efficiency higher than 99\% and to keep hit noise occupancy lower than 0.1\% throughout the end of their lifetime. This work examines the impact of CN on the tracking performance and assesses whether the modules can operate within specifications when affected by CN. We tested both irradiated and non-irradiated short strip modules impacted by CN using a 5 GeV electron beam. The tracking performance was evaluated for each individual strip as well as for the overall module.

From the performance studies of single strips, the results indicated that CN narrows the width of the operating window. A statistical analysis of three regions (normal region, low CN region, and high CN region) revealed that when high CN is present, a small percentage of strips may fail to meet detector requirements before irradiation. However, over half of the strips fail after irradiation at the expected end-of-life fluence. The fraction of strips without an operating window due to CN at any point in the lifetime can be predicted according to the measured~$\mathrm{Q_c^{no}}$ and its expected~$\mathrm{Q_{50}}$. From the global performance studies, the non-irradiated module can be operated within specification by increasing the threshold to 1 fC; however, it is impossible to operate the irradiated module within specification. The studies indicated that the modules affected by CN may be used at low fluence spots of the barrel region. Moreover, the studies indicated that if the hit noise occupancy caused by CN can be smaller than 1\% at threshold higher than 0.45 fC, approximately 60\% of strips can fulfill the operating requirements by the end of the detector's lifetime, making it highly likely that the module will meet the requirements regarding the global performance. 

After two years of studies, as of the time of writing, the so-called "interposer" solution is being pursued by the collaboration. The solution involves redesigning the module by adding layers of soft adhesive and a kapton separator between the power board and the sensor so that the sensor is shielded from the vibration of the power board. Initial test beam results indicate that the modules with the interposed layer meet the specified requirements.

\FloatBarrier
\section*{Acknowledgments}

The authors would like to thank the crew at the IRRAD facility at CERN for their help with the irradiations.

The measurements leading to these results have been performed at the Test Beam Facility at DESY Hamburg (Germany), a member of the Helmholtz Association (HGF).

This work was supported by the European Structural and Investment Funds and the Ministry of Education, Youth and Sports of the Czech Republic via projects LM2023040 CERN-CZ and FORTE - CZ.02.01.01/00/22\_008/0004632.

\appendix
\section{Noise Estimation}
\label{sec:appendixA}
\begin{figure*}[]
\centering
	\begin{subfigure}{0.50\linewidth}
		\includegraphics[width=\linewidth]{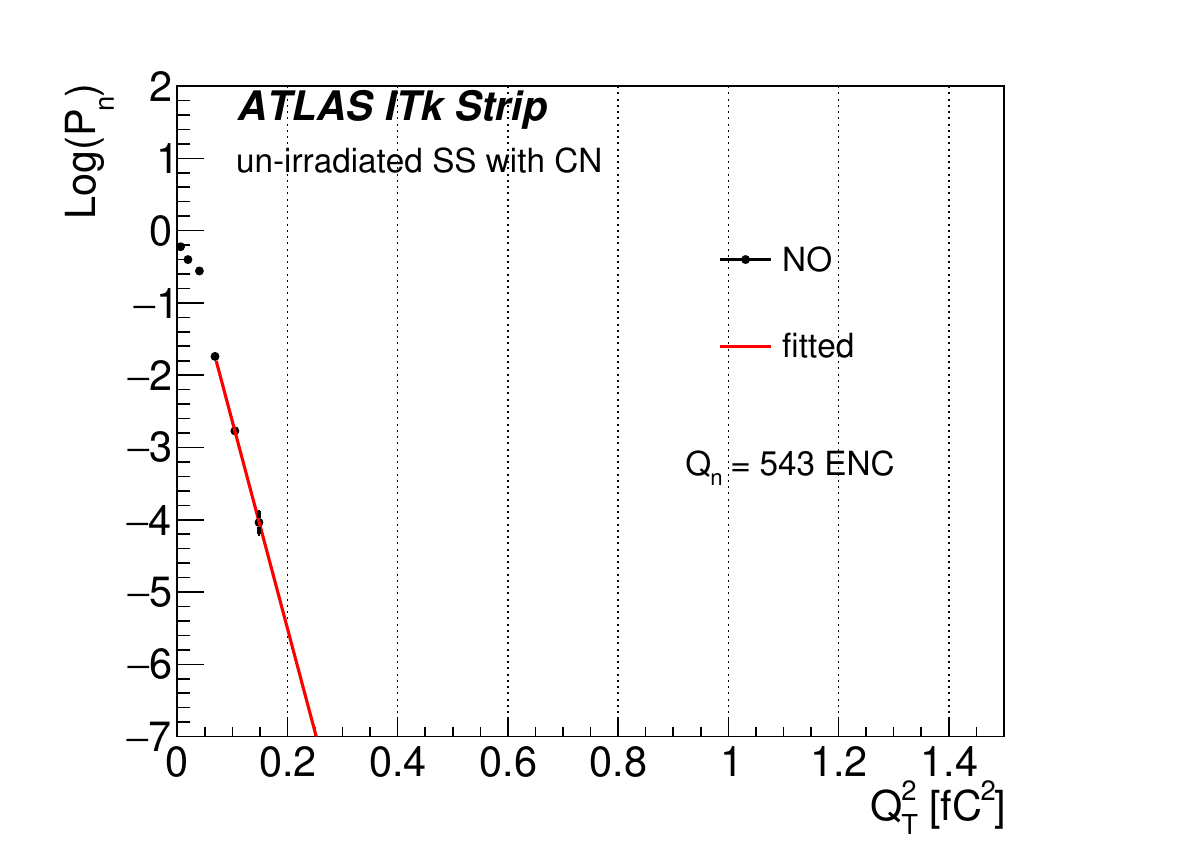}
		\caption{Normal strip}
        \label{fig:no-normal}
	\end{subfigure}	
	\begin{subfigure}{0.50\linewidth}
		\includegraphics[width=\linewidth]{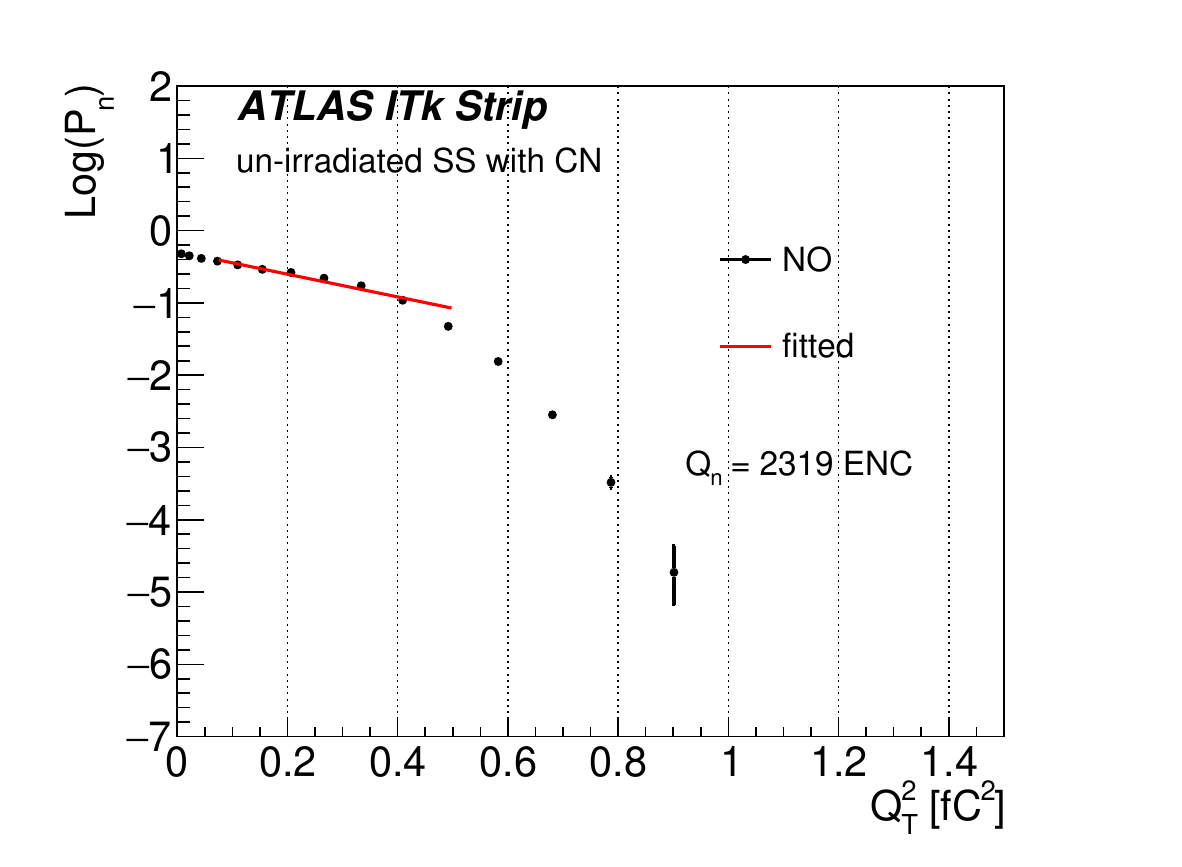}
		\caption{CN strip}
        \label{fig:no-coldnoise-small}
	\end{subfigure}	
    \begin{subfigure}{0.50\linewidth}
		\includegraphics[width=\linewidth]{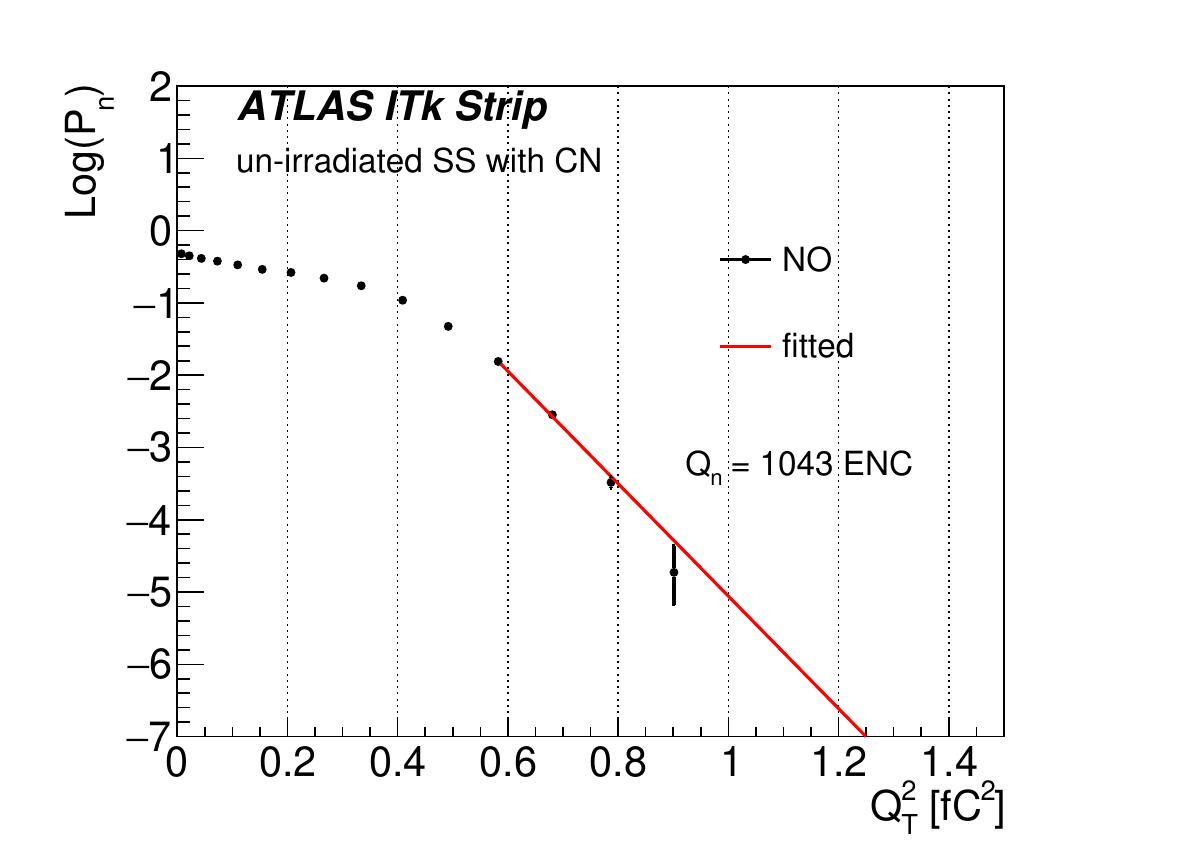}
		\caption{CN strip}
        \label{fig:no-coldnoise-big}
	\end{subfigure}	
\caption{The logarithm of $\mathrm{P_n}$ as a function of $\mathrm{Q_T^2}$ for normal and CN strip from s20 of the non-irradiated module. The logarithm of $\mathrm{P_n}$ can be fitted with a linear function of $\mathrm{Q_T^2}$ for a normal strip. For Cold Nosie strips, there are two distinct slopes that need fitting. The red line represents the linear fitted curve of the distribution. The fitting regions are specified as follows:~$\mathrm{Q_T^2 > 0.05 \ fC^2}$ in (a),~$\mathrm{0.05 < Q_T^2 < 0.4 \ fC^2}$ in (b) and~$\mathrm{P_n < 10^{-2}}$ in (c).~$\mathrm{Q_n}$ is the equivalent noise charge estimated from the fitting curves.}
\label{fig:pedestal-no-curve}
\end{figure*}

Figure~\ref{fig:no-normal} illustrates the logarithm of $\mathrm{P_n}$ as a function of $\mathrm{Q_T^2}$ for a normal strip from s20. For~$\mathrm{Q_T^2 < 0.05 \ fC^2}$, the common mode noise is dominant. For~$\mathrm{Q_T^2 > 0.05 \  fC^2}$, the logarithm of $\mathrm{P_n}$ and $\mathrm{Q_T^2}$ shows a linear correlation and the slope is fitted.~$\mathrm{Q_n}$ is estimated to be 543 equivalent noise charge (ENC). Figure~\ref{fig:no-coldnoise-small} and~\ref{fig:no-coldnoise-big} present the logarithm of $\mathrm{P_n}$ as a function of $\mathrm{Q_T^2}$ for a CN strip from s20. This distribution exhibits more complex features. The estimation of $\mathrm{Q_n}$ depends strongly on the fitting range of $\mathrm{Q_T^2}$. Figure~\ref{fig:no-coldnoise-small} shows the fit result for the range of~$\mathrm{0.05 < Q_T^2 < 0.4 \ fC^2}$. Figure~\ref{fig:no-coldnoise-big} displays the fit results for the region where~$\mathrm{P_n < 10^{-2}}$. Because CN is not Gaussian behavior noise, the estimated $\mathrm{Q_n}$ from the two regions diverge, corresponding 2319 ENC and 1043 ENC, respectively. 

\begin{figure*} []
\centering
	\begin{subfigure}{0.49\linewidth}
		\includegraphics[width=\linewidth]{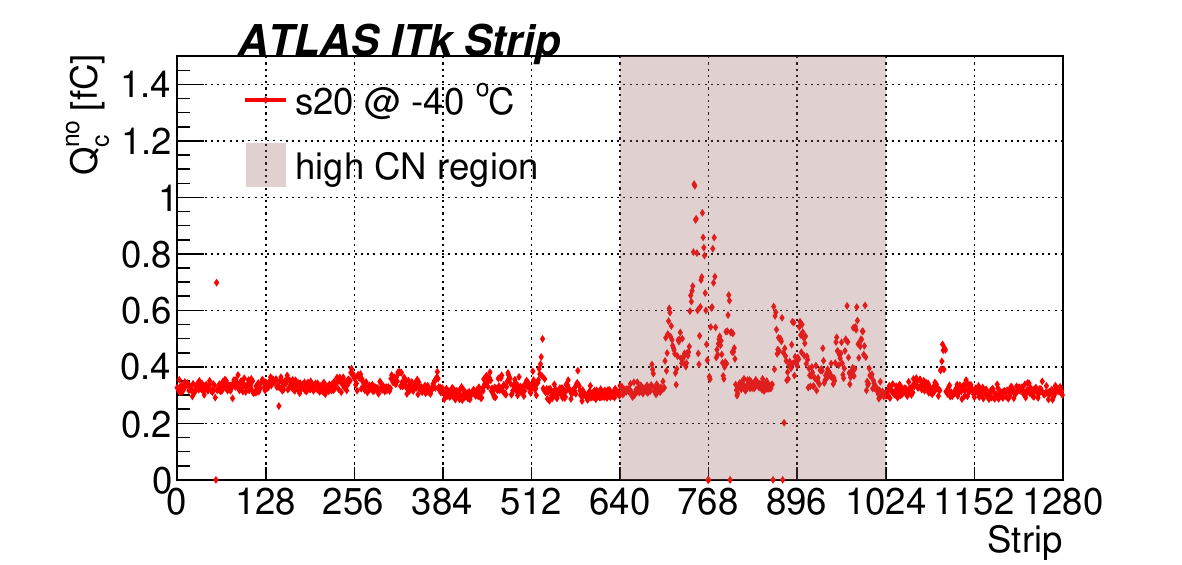}
		\caption{s20,~$\mathrm{\overline{Q_c^{no}} = 0.33 \pm 0.03 \ fC}$}
        \label{fig:ps20}
	\end{subfigure}	
	\begin{subfigure}{0.49\linewidth}
		\includegraphics[width=\linewidth]{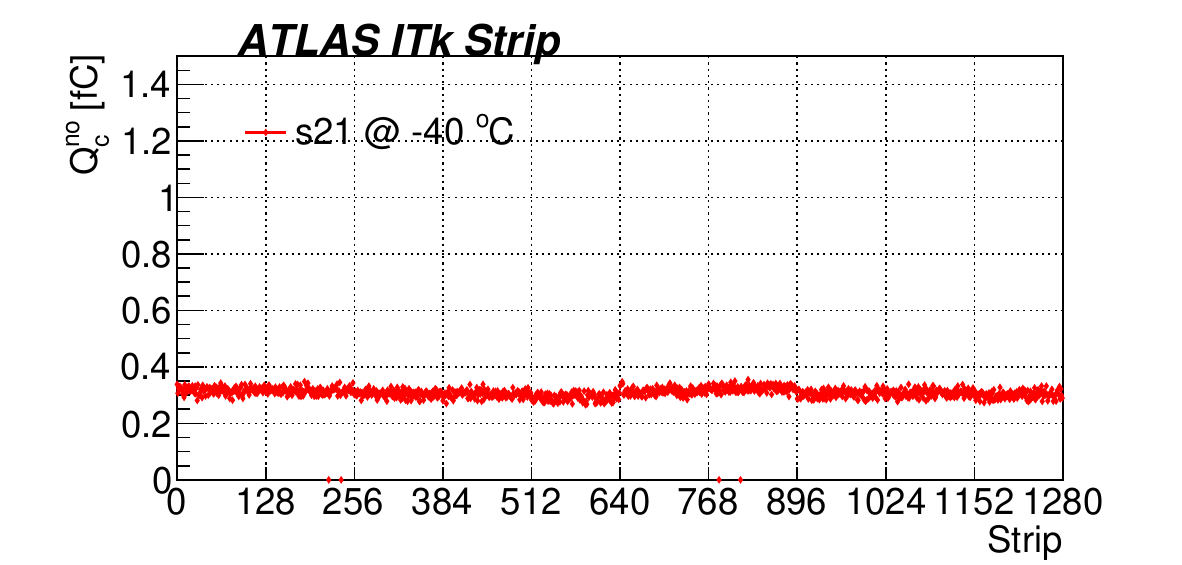}
		\caption{s21,~$\mathrm{\overline{Q_c^{no}} = 0.31 \pm 0.02 \ fC}$}
        \label{fig:ps21}
	\end{subfigure}	
    \begin{subfigure}{0.49\linewidth}
		\includegraphics[width=\linewidth]{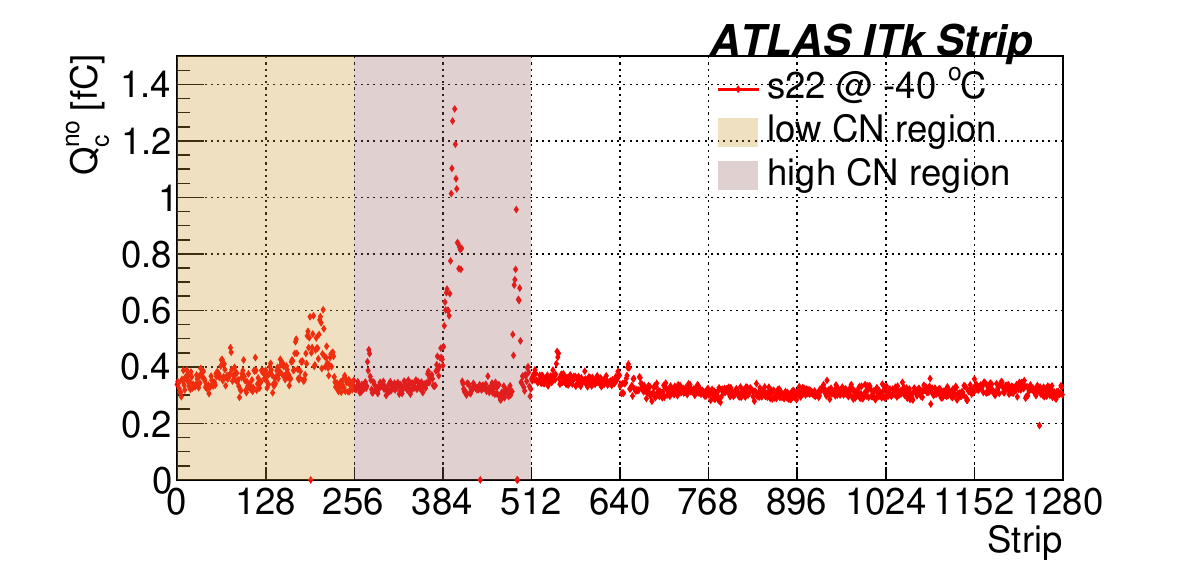}
		\caption{s22,~$\mathrm{\overline{Q_c^{no}} = 0.32 \pm 0.02 \ fC}$}
        \label{fig:ps22}
	\end{subfigure}	
	\begin{subfigure}{0.49\linewidth}
		\includegraphics[width=\linewidth]{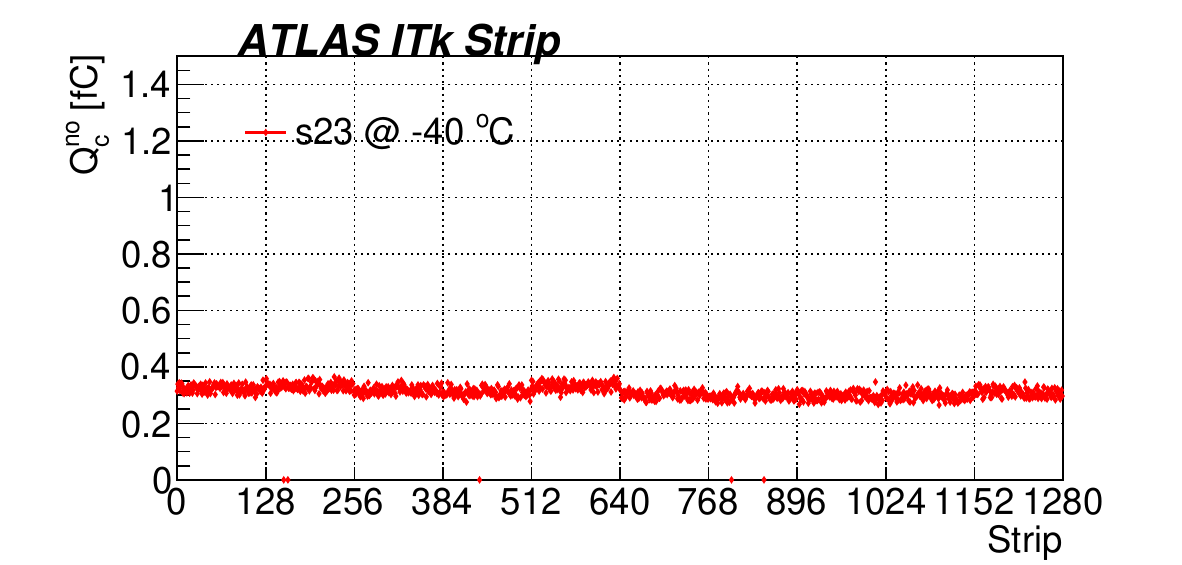}
		\caption{s23,~$\mathrm{\overline{Q_c^{no}} = 0.31 \pm 0.03 \ fC}$}
        \label{fig:ps23}
	\end{subfigure}	
\caption{~$\mathrm{Q_c^{no}}$ of the non-irradiated short strip module. The noise hits are taken using an external trigger rate of 3 kHz. The low and high CN regions are highlighted. ~$\mathrm{\overline{Q_c^{no}}}$ is the average of ~$\mathrm{Q_c^{no}}$ of the strips not in the CN regions.}
\label{fig:pedestal-noise-unirr}
\end{figure*}

\begin{figure*}[]
\centering
	\begin{subfigure}{0.49\linewidth}
		\includegraphics[width=\linewidth]{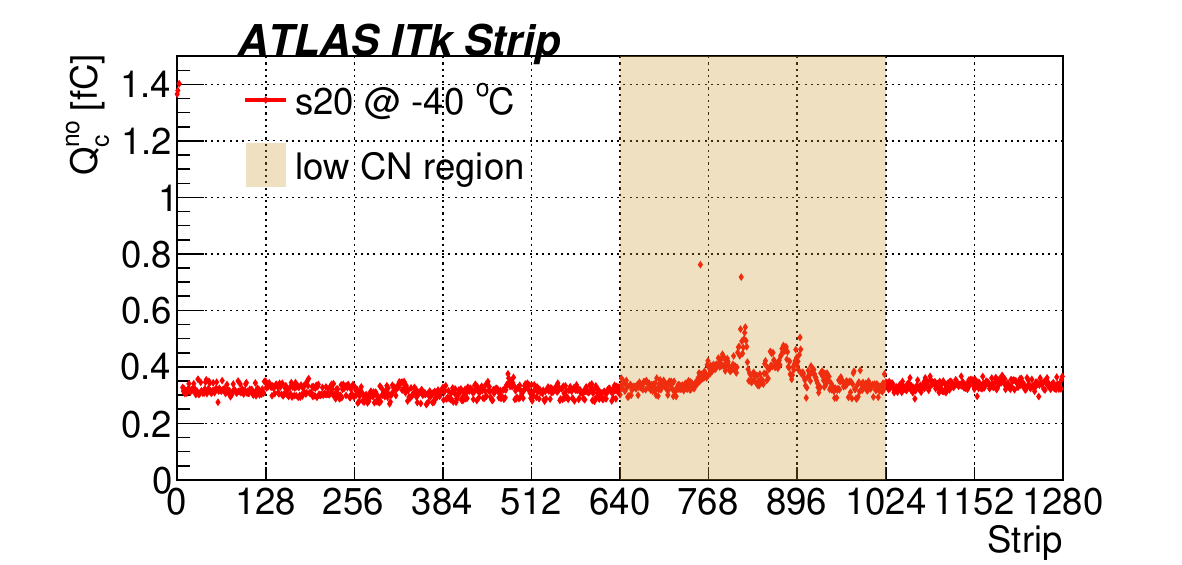}
		\caption{s20,~$\mathrm{\overline{Q_c^{no}} = 0.32 \pm 0.02 \ fC}$}
        \label{fig:ps20}
	\end{subfigure}	
	\begin{subfigure}{0.49\linewidth}
		\includegraphics[width=\linewidth]{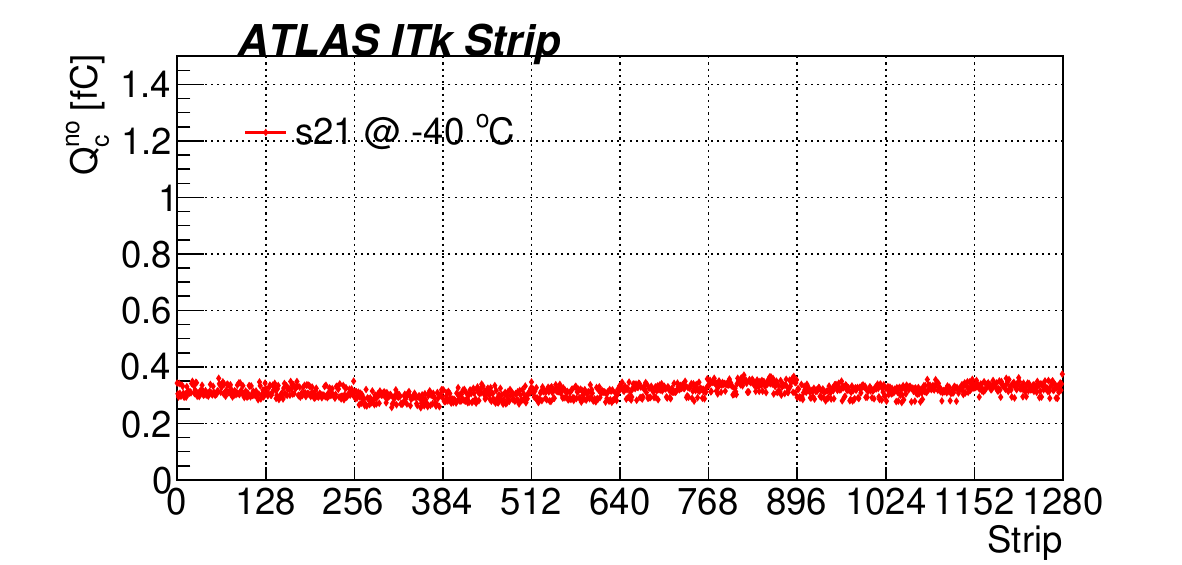}
		\caption{s21,~$\mathrm{\overline{Q_c^{no}} = 0.31 \pm 0.02 \ fC}$}
        \label{fig:ps21}
	\end{subfigure}	
    \begin{subfigure}{0.49\linewidth}
		\includegraphics[width=\linewidth]{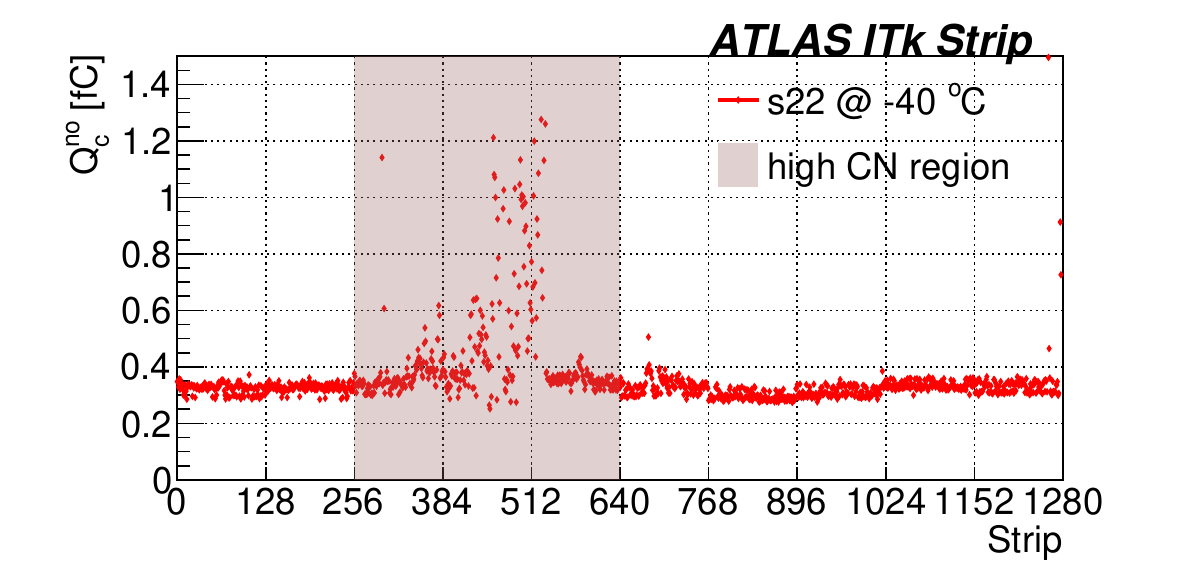}
		\caption{s22,~$\mathrm{\overline{Q_c^{no}} = 0.32 \pm 0.06 \ fC}$}
        \label{fig:ps22}
	\end{subfigure}	
	\begin{subfigure}{0.49\linewidth}
		\includegraphics[width=\linewidth]{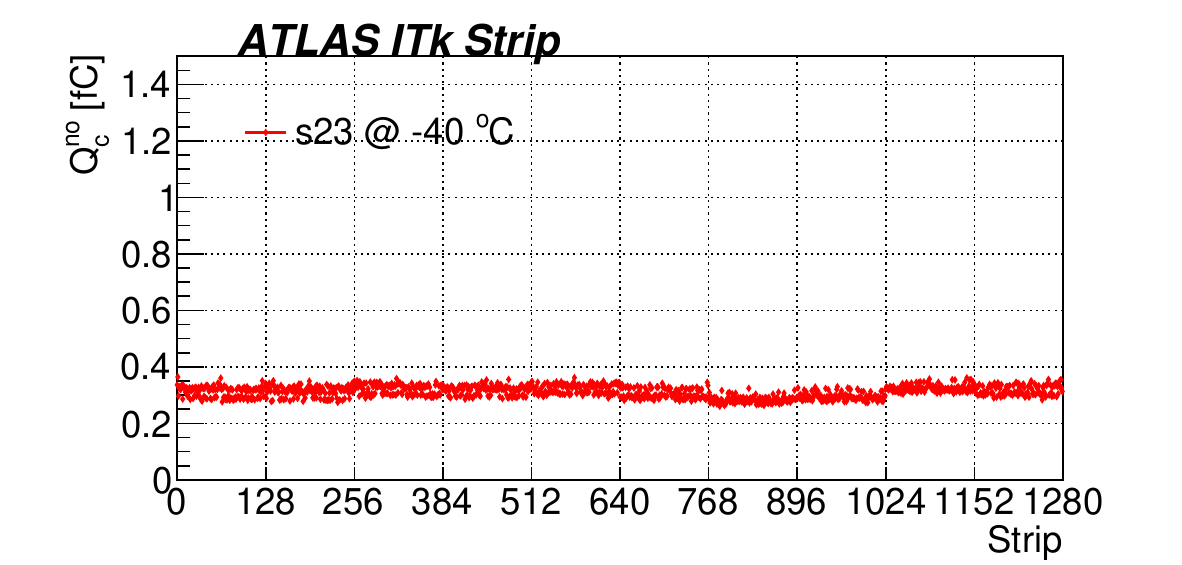}
		\caption{s23,~$\mathrm{\overline{Q_c^{no}} = 0.31 \pm 0.02 \ fC}$}
        \label{fig:ps23}
	\end{subfigure}	
\caption{~$\mathrm{Q_c^{no}}$of the sensor irradiated short strip module. The noise hits are taken using an external trigger rate of 3 kHz. The low and high CN regions are highlighted. ~$\mathrm{\overline{Q_c^{no}}}$~is the average of~$\mathrm{Q_c^{no}}$ of the strips not in the CN regions.}
\label{fig:pedestal-noise-irr}
\end{figure*}
\section{Median Collected Charge}
\label{app:q50}
\begin{figure*}[]
\centering
	\begin{subfigure}{0.49\linewidth}
		\includegraphics[width=\linewidth]{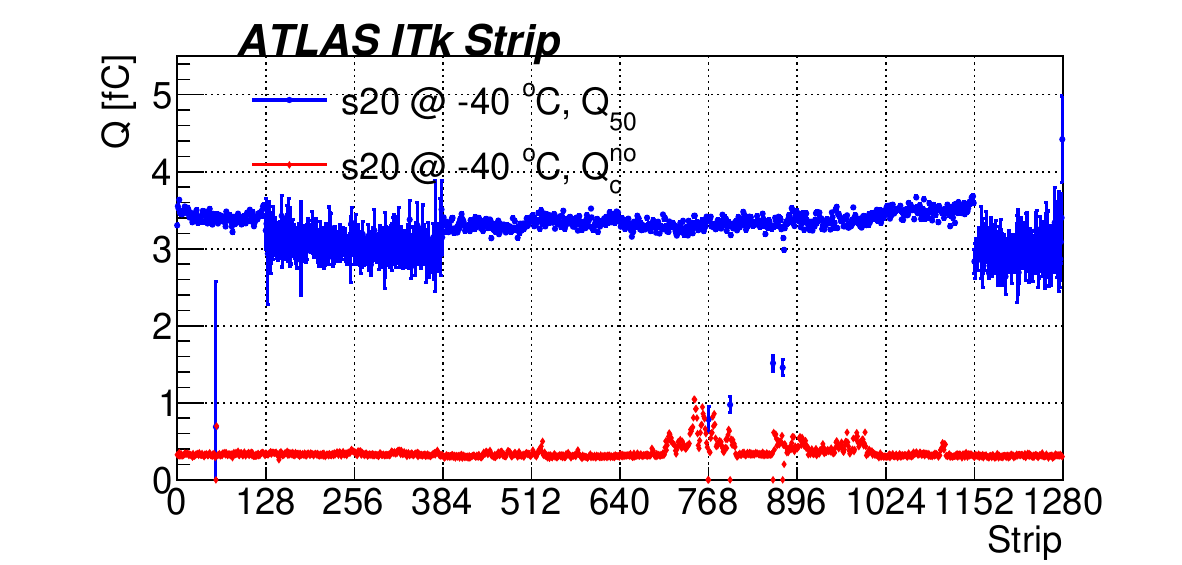}
		\caption{s20}
        \label{fig:q50s20}
	\end{subfigure}	
	\begin{subfigure}{0.49\linewidth}
		\includegraphics[width=\linewidth]{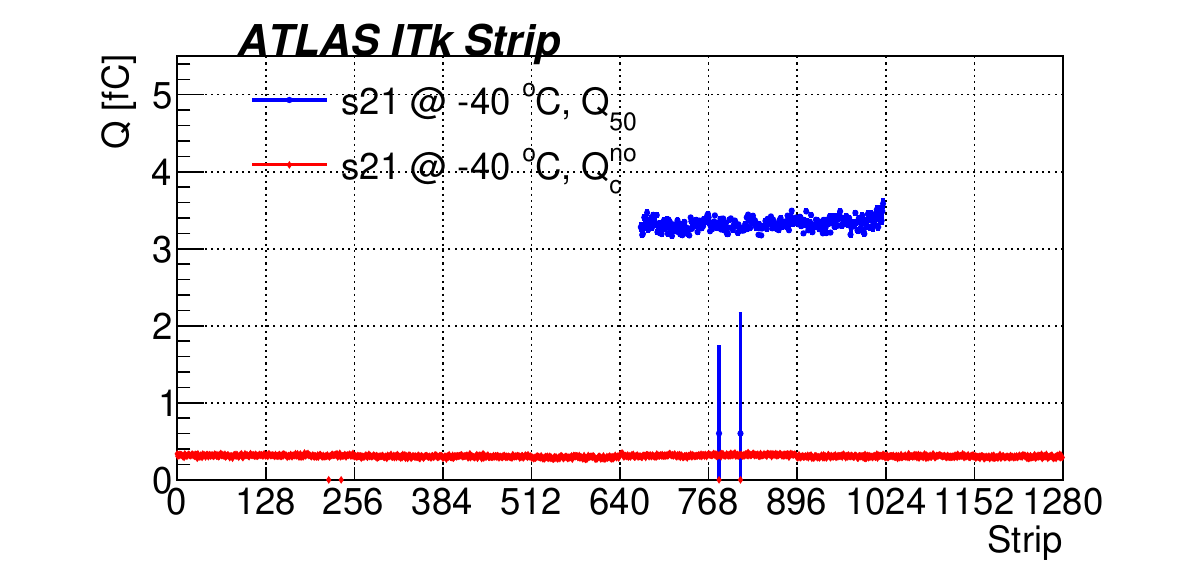}
		\caption{s21}
        \label{fig:q50s21}
	\end{subfigure}	
    \begin{subfigure}{0.49\linewidth}
		\includegraphics[width=\linewidth]{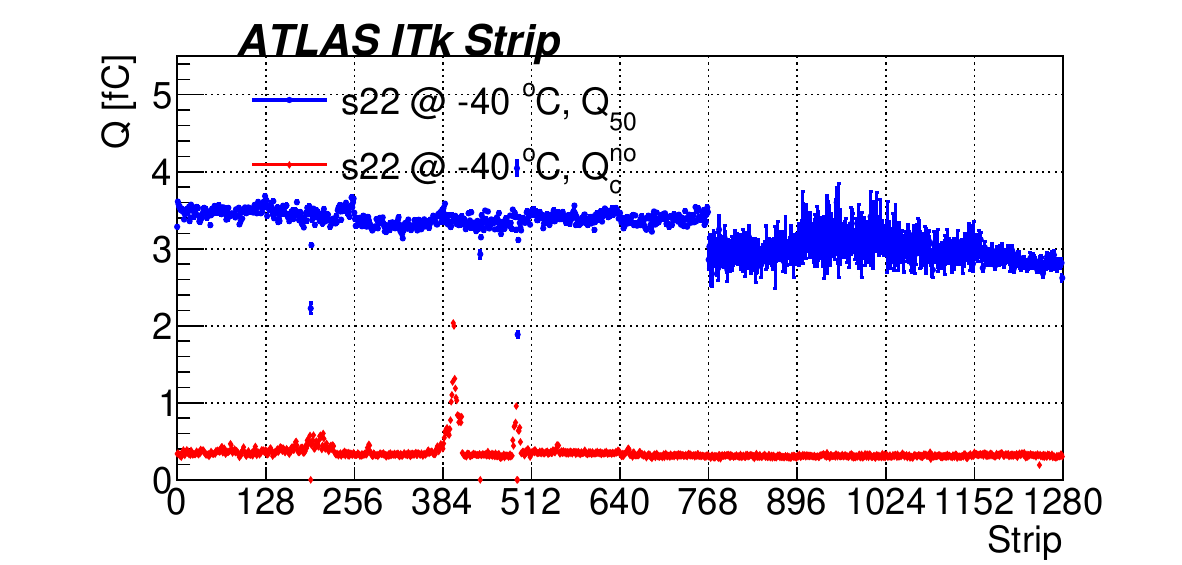}
		\caption{s22}
        \label{fig:q50s22}
	\end{subfigure}	
	\begin{subfigure}{0.49\linewidth}
		\includegraphics[width=\linewidth]{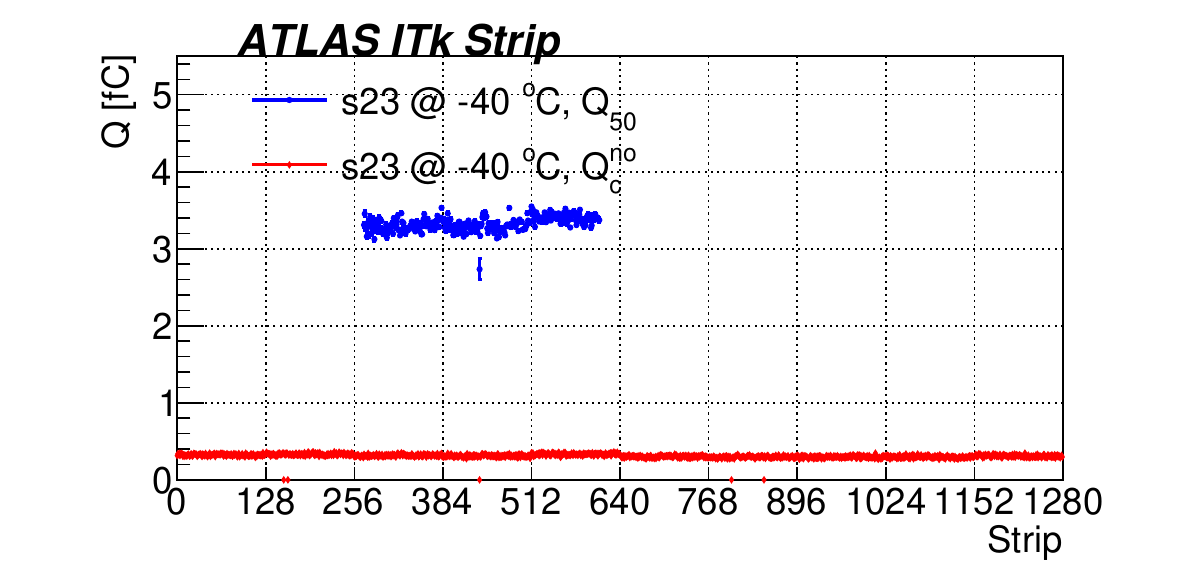}
		\caption{s23}
        \label{fig:q50s23}
	\end{subfigure}	
\caption{~$\mathrm{Q_{50}}$ and~$\mathrm{Q_c^{no}}$ for each strip tested with the beam in streams of the non-irradiated module. The error bar for~$\mathrm{Q_{50}}$ represents the fitted error. The results for s20 and s22 are combined through an efficiency analysis of threshold scans conducted at seven different positions, specifically for the following strip ranges: 0-127, 128-383, 384-511, 512-767, 768-895, 896-1151, and 1152-1279. The results for s21 are obtained from the efficiency analysis of a threshold scan for strips 670-1020, while the results for s23 come from the efficiency analysis of a threshold scan for strips 260-610. The average of~$\mathrm{Q_{50}}$ is summarised in Table~\ref{tab:q50}.}
\label{fig:unirr-q50}
\end{figure*}

\begin{figure*}[]
\centering
	\begin{subfigure}{0.49\linewidth}
		\includegraphics[width=\linewidth]{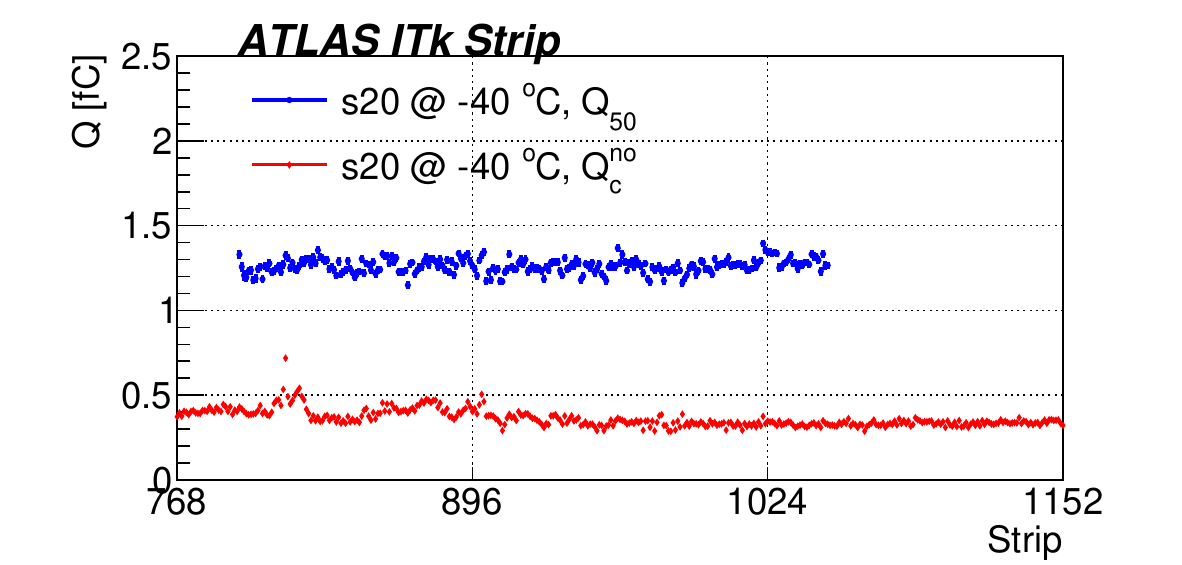}
		\caption{s20}
        \label{fig:irr-q50s20}
	\end{subfigure}	
    \begin{subfigure}{0.49\linewidth}
		\includegraphics[width=\linewidth]{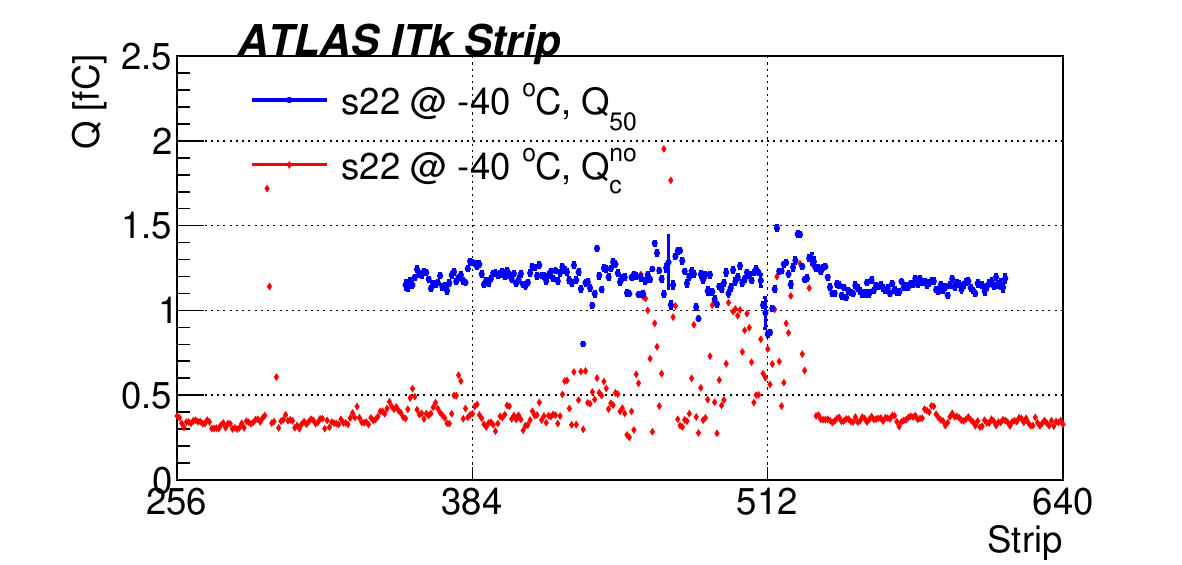}
		\caption{s22}
        \label{fig:irr-q50s22}
	\end{subfigure}	
\caption{$\mathrm{Q_{50}}$ and~$\mathrm{Q_c^{no}}$ for each strips tested in s20 and s22 of the irradiated module. The error bar of~$\mathrm{Q_{50}}$ represents the fitted error. The results for s20 are derived from the efficiency analysis of the threshold scan for strips 795 to 1050. The results for s22 come from the efficiency analysis of the threshold scan for strips 355 to 615.  The average of~$\mathrm{Q_{50}}$ is summarised in Table~\ref{tab:q50}.}
\label{fig:irr-q50}
\end{figure*}

The median collected charge for each strip in all streams of the non-irradiated module are displayed in Figure~\ref{fig:unirr-q50}. The value of~$\mathrm{Q_{50}}$ is plotted alongside~$\mathrm{Q_c^{no}}$ to illustrate the impact of CN. Strips 128-383 and 1152-1279 in s20 exhibit larger fitted error bars compared to other strips. This is because thresholds higher than 2 fC at these positions were not scanned, which complicates the estimation of~$\mathrm{Q_{50}}$ from the fit. A similar situation is observed for strips 768-1279 in s22. A few dead strips are present in the module, showing~$\mathrm{Q_c^{no}}$ of 0. Since only charges shared to their active neighbours can be measured,~$\mathrm{Q_{50}}$ is significantly reduced. The strips mentioned above and the dead strips are excluded when calculating the average of~$\mathrm{Q_{50}}$ among strips. Figure~\ref{fig:irr-q50} illustrates the median collected charge of strips in the CN region of the irradiated module, together with~$\mathrm{Q_c^{no}}$. For strips in s20, the median collected charge is unaffected by CN. Although in s22, fluctuations in the median collected charge are observed among strips 440-540 within the CN region, the average~$\mathrm{Q_{50}}$ of those strips is consistent with the median collected charge estimated from normal strips in this region, for example strip 550-610.

\section{Operating Window}
\label{app:opw}
\begin{figure*}[]
\centering
	\begin{subfigure}{0.49\linewidth}
		\includegraphics[width=\linewidth]{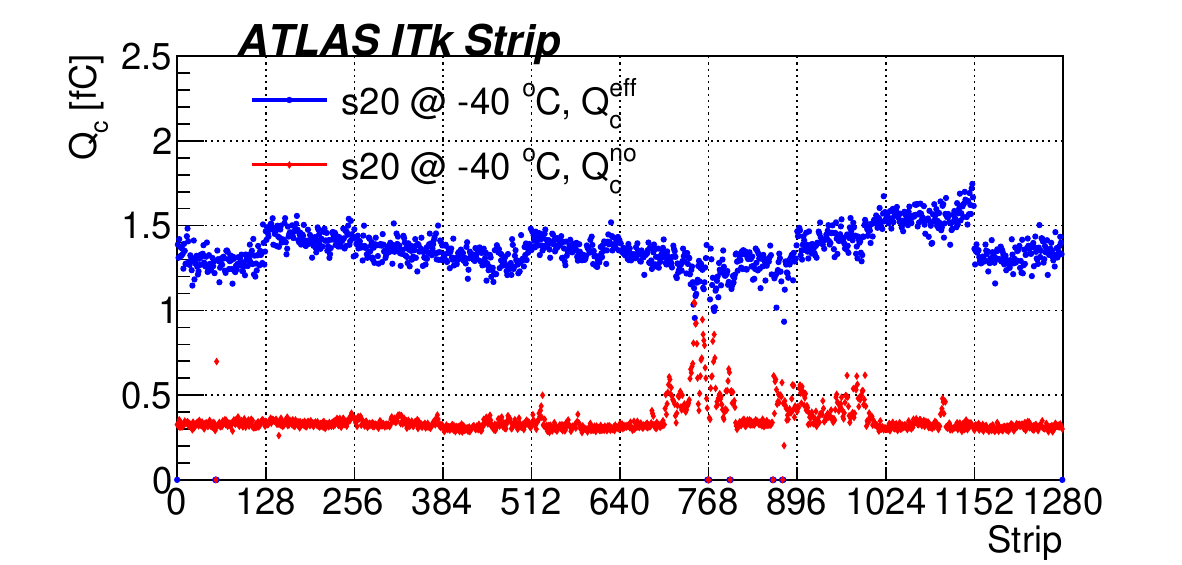}
		\caption{s20}
        \label{fig:qeffs20}
	\end{subfigure}	
	\begin{subfigure}{0.49\linewidth}
		\includegraphics[width=\linewidth]{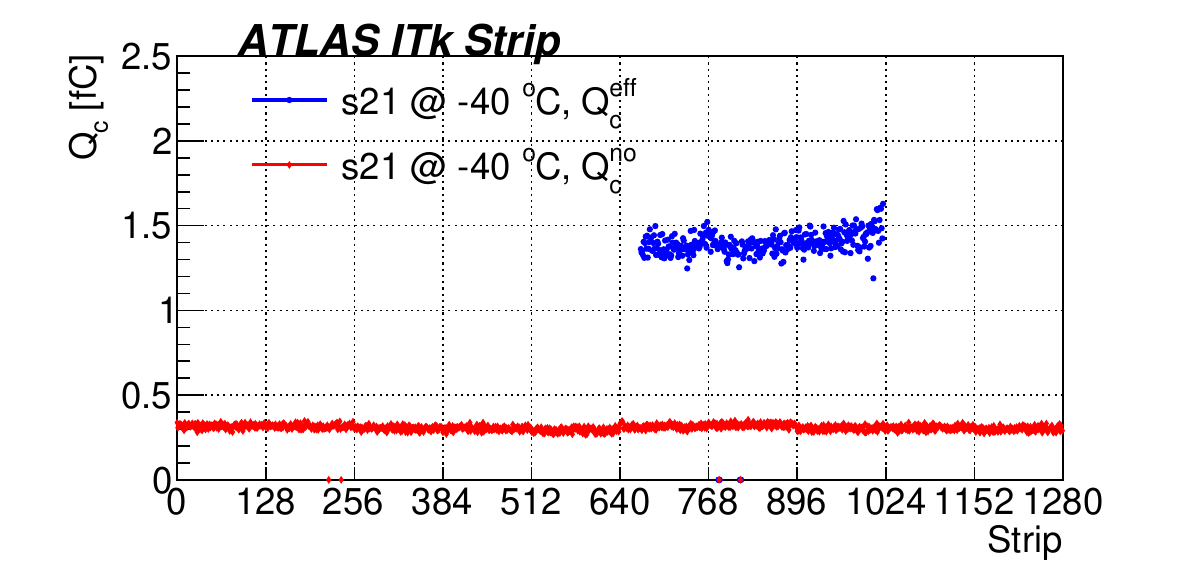}
		\caption{s21}
        \label{fig:qeffs21}
	\end{subfigure}	
    \begin{subfigure}{0.49\linewidth}
		\includegraphics[width=\linewidth]{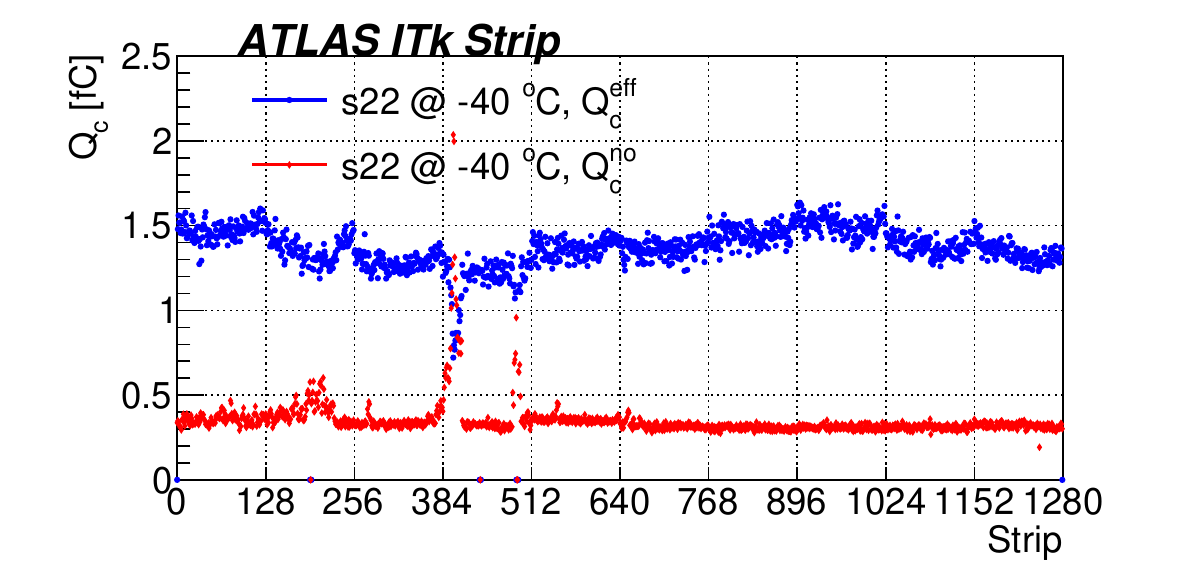}
		\caption{s22}
        \label{fig:qeffs22}
	\end{subfigure}	
	\begin{subfigure}{0.49\linewidth}
		\includegraphics[width=\linewidth]{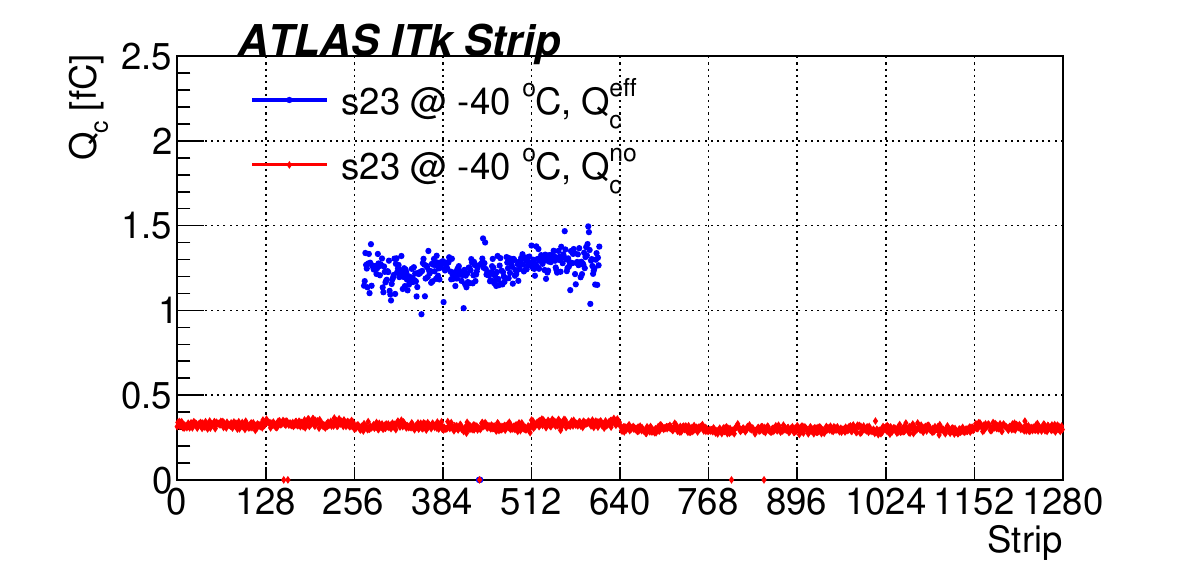}
		\caption{s23}
        \label{fig:qeffs23}
	\end{subfigure}	
\caption{$\mathrm{Q_c^{eff}}$ and~$\mathrm{Q_c^{no}}$ for strips in all streams of the non-irradiated module. If the maximum efficiency is lower than~99\%, a value of 0 is assigned to ~$\mathrm{Q_c^{eff}}$.}
\label{fig:unirr-opw}
\end{figure*}

\begin{figure*}[]
    \centering
    \begin{subfigure}{0.49\linewidth}
        \includegraphics[width=\linewidth]{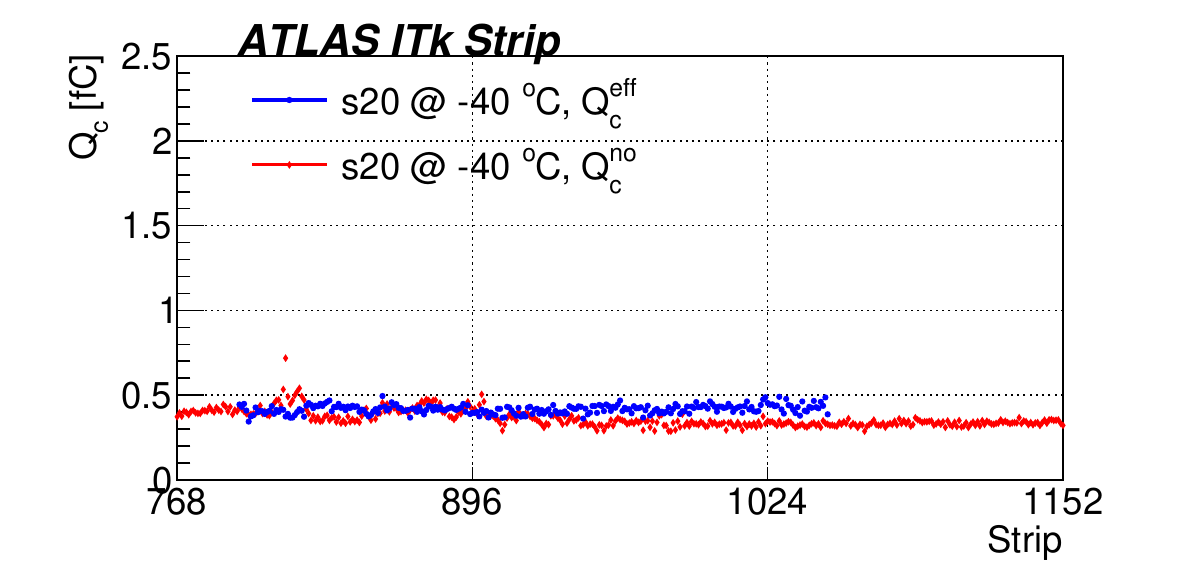}
        \caption{s20}
        \label{fig:irr-opws20}
    \end{subfigure}
    \begin{subfigure}{0.49\linewidth}
        \includegraphics[width=\linewidth]{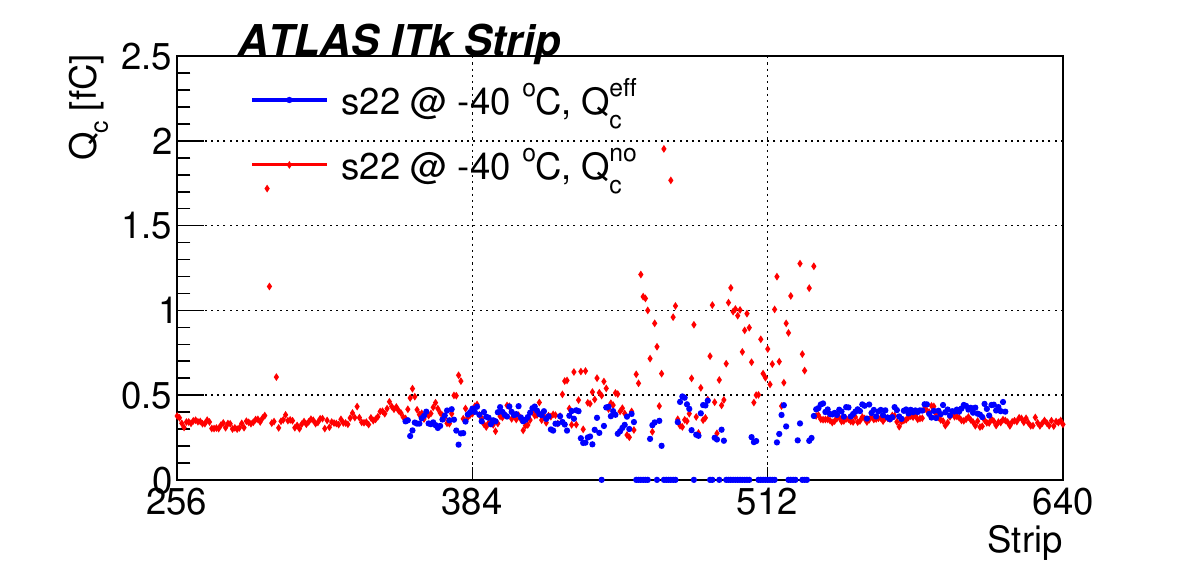}
        \caption{s22}
        \label{fig:irr-opws21}
    \end{subfigure}
    \caption{$\mathrm{Q_c^{eff}}$ and~$\mathrm{Q_c^{no}}$ for strips in s20 and s22 of the irradiated module. If the maximum efficiency is lower than~99\%, the value of 0 is assigned to ~$\mathrm{Q_c^{eff}}$.}
    \label{fig:irr-opw}
\end{figure*}

Figure~\ref{fig:unirr-opw} illustrates~$\mathrm{Q_c^{eff}}$ for each strip across all streams of the non-irradiated module. In s21 and s23,~$\mathrm{Q_c^{eff}}$ remains consistent among the strips. However,~$\mathrm{Q_c^{eff}}$ varies among chips in s20 and s22 due to different beam conditions during data collection at different positions, which affects the approximation of~$\mathrm{Q_c^{eff}}$ as pointed in Section~\ref{sec:eff_ana}. Despite this variation,~$\mathrm{Q_c^{eff}}$ for strips not in the CN regions is larger than 1 fC. c§Figure~\ref{fig:irr-opw} shows ~$\mathrm{Q_c^{eff}}$ for s20 and s22 of the irradiated module. ~$\mathrm{Q_c^{eff}}$ drops to below 0.5 fC and distributes uniformly among strips.

\bibliography{mybibfile.bib}

@techreport{CERN-LHCC-2017-005,
      collaboration = "ATLAS",
      title         = "{Technical Design Report for the ATLAS Inner Tracker Strip
                       Detector}",
      institution   = "CERN",
      reportNumber  = "CERN-LHCC-2017-005, ATLAS-TDR-025",
      address       = "Geneva",
      year          = "2017",
      url           = "https://cds.cern.ch/record/2257755",
}

@article{Dyckes_2024,
doi = {10.1088/1748-0221/19/04/C04058},
year = {2024},
month = {apr},
publisher = {IOP Publishing},
volume = {19},
number = {04},
pages = {C04058},
author = "{Dyckes, G.I. and Kurth, M.G. and on behalf of the ITk Strip collaboration}",
title = "{How the discovery of Cold Noise delayed the production of ATLAS ITk strip tracker modules by a year}",
journal = {Journal of Instrumentation}
}

@article{Arling:2023pio,
    author = "Arling, J. -H. and others",
    title = "{Test beam measurement of ATLAS ITk Short Strip module at warm and cold operational temperature}",
    eprint = "2302.10950",
    archivePrefix = "arXiv",
    primaryClass = "hep-ex",
    doi = "10.1088/1748-0221/18/03/P03015",
    journal = "JINST",
    volume = "18",
    number = "03",
    pages = "P03015",
    year = "2023"
}

@article{Arling_2025,
doi = {10.1088/1748-0221/20/07/P07033},
year = {2025},
month = {jul},
publisher = {IOP Publishing},
volume = {20},
number = {07},
pages = {P07033},
author = {Arling, J.-H. and others},
title = "{Test beam measurements and computer simulations of the ATLAS ITk R2 silicon strip detector}",
journal = {Journal of Instrumentation}
}

@article{Lu_2017,
doi = {10.1088/1748-0221/12/04/C04017},
year = {2017},
month = {apr},
publisher = {},
volume = {12},
number = {04},
pages = {C04017},
author = {Lu, W. and others},
title = "{Development of the ABCStar front-end chip for the ATLAS silicon  strip upgrade}",
journal = {Journal of Instrumentation}
}

@article{Cormier_2021,
doi = {10.1088/1748-0221/16/07/P07061},
url = {https://doi.org/10.1088/1748-0221/16/07/P07061},
year = {2021},
month = {jul},
publisher = {IOP Publishing},
volume = {16},
number = {07},
pages = {P07061},
author = {Cormier, K.J.R. and others},
title = {Development of the front end amplifier circuit for the ATLAS ITk silicon strip detector},
journal = {Journal of Instrumentation}
}

@inproceedings{ravotti:hal-03011049,
  TITLE = "{{The IRRAD Proton Irradiation Facility Control, Data Management and Beam Diagnostic Systems: An Outlook of the Major Upgrades Beyond the CERN Long Shutdown 2}}",
  AUTHOR = {Ravotti, Federico and others},
  BOOKTITLE = {{17th International Conference on Accelerator and Large Experimental Physics Control Systems}},
  ADDRESS = {New York, United States},
  PAGES = {WEPHA127},
  YEAR = {2019},
  MONTH = Oct,
  DOI = {10.18429/JACoW-ICALEPCS2019-WEPHA127},
  HAL_ID = {hal-03011049},
  HAL_VERSION = {v1},
}

@article{Unno_2023,
doi = {10.1088/1748-0221/18/03/T03008},
year = {2023},
month = {mar},
publisher = {IOP Publishing},
volume = {18},
number = {03},
pages = {T03008},
author = {Unno, Y. and others},
title = "{Specifications and pre-production of n+-in-p large-format strip sensors fabricated in 6-inch silicon wafers, ATLAS18, for the Inner Tracker of the ATLAS Detector for High-Luminosity Large Hadron Collider}",
journal = {Journal of Instrumentation}
}

@book{Spieler:2005si,
    author = "Spieler, Helmuth",
    title = "{Semiconductor Detector Systems}",
    isbn = "978-0-19-852784-8",
    publisher = "Oxford University Press",
    address = "Oxford",
    series = "Semiconductor Science and Technology",
    volume = "v.12",
    year = "2005"
}

@article{DIENER2019265,
title = "{The DESY II test beam facility}",
journal = {Nuclear Instruments and Methods in Physics Research Section A: Accelerators, Spectrometers, Detectors and Associated Equipment},
volume = {922},
pages = {265-286},
year = {2019},
issn = {0168-9002},
doi = {10.1016/j.nima.2018.11.133},
author = {R. Diener and others}
}

@article{Liu:2023uas,
    author = "Liu, Yi and others",
    title = "{ADENIUM \textemdash{} A demonstrator for a next-generation beam telescope at DESY}",
    eprint = "2301.05909",
    archivePrefix = "arXiv",
    primaryClass = "physics.ins-det",
    reportNumber = "PUBDB-2022-07438",
    doi = "10.1088/1748-0221/18/06/P06025",
    journal = "JINST",
    volume = "18",
    number = "06",
    pages = "P06025",
    year = "2023"
}

@article{Baesso:2019smg,
    author = "Baesso, P. and Cussans, D. and Goldstein, J.",
    title = "{The AIDA-2020 TLU: a flexible trigger logic unit for test beam facilities}",
    eprint = "2005.00310",
    archivePrefix = "arXiv",
    primaryClass = "physics.ins-det",
    doi = "10.1088/1748-0221/14/09/P09019",
    journal = "JINST",
    volume = "14",
    number = "09",
    pages = "P09019",
    year = "2019"
}

@article{Dannheim:2020jlk,
    author = "Dannheim, Dominik and others",
    title = "{Corryvreckan: A Modular 4D Track Reconstruction and Analysis Software for Test Beam Data}",
    eprint = "2011.12730",
    archivePrefix = "arXiv",
    primaryClass = "physics.ins-det",
    reportNumber = "CLICdp-Pub-2020-005, DESY-20-210",
    doi = "10.1088/1748-0221/16/03/P03008",
    journal = "JINST",
    volume = "16",
    number = "03",
    pages = "P03008",
    year = "2021"
}

@article{HUTH2025170720,
title = "{TelePix2: Full scale fast region of interest trigger and timing for the EUDET-style telescopes at the DESY II test beam facility}",
journal = {Nuclear Instruments and Methods in Physics Research Section A: Accelerators, Spectrometers, Detectors and Associated Equipment},
volume = {1080},
pages = {170720},
year = {2025},
issn = {0168-9002},
doi = {10.1016/j.nima.2025.170720},
author = {L. Huth and others}
}

@article{Liu:2019wim,
    author = "Liu, Y. and others",
    title = "{EUDAQ2\textemdash{}A flexible data acquisition software framework for common test beams}",
    eprint = "1907.10600",
    archivePrefix = "arXiv",
    primaryClass = "physics.ins-det",
    doi = "10.1088/1748-0221/14/10/P10033",
    journal = "JINST",
    volume = "14",
    number = "10",
    pages = "P10033",
    year = "2019"
}

@article{eff_fit-ref,
title = "{Beam tests of ATLAS SCT silicon strip detector modules}",
journal = {Nuclear Instruments and Methods in Physics Research Section A: Accelerators, Spectrometers, Detectors and Associated Equipment},
volume = {538},
number = {1},
pages = {384-407},
year = {2005},
issn = {0168-9002},
doi = {10.1016/j.nima.2004.08.133},
author = {F. Campabadal and others}
}
\end{document}